\documentclass[aps,prl,twocolumn,amsmath,amssymb,showpacs,superscriptaddress,notitlepage,longbibliography,floatfix]{revtex4-2}
\usepackage{array}
\usepackage{graphicx}
\usepackage{subfigure}
\usepackage{dcolumn}
\usepackage{bm}
\usepackage{amsfonts}
\usepackage{mathrsfs}
\usepackage{amssymb}
\usepackage{amsmath}
\usepackage{color}
\usepackage{chngcntr}
\usepackage{stmaryrd}
\usepackage{xcolor}
\usepackage[colorlinks=true,breaklinks=true,linkcolor=blue,anchorcolor=blue,citecolor=blue,urlcolor=blue]{hyperref}
\usepackage{booktabs}
\usepackage{multirow}
\usepackage{tcolorbox}
\usepackage{titlesec}
\date{\today}
\titleformat{\section}[block]{\normalfont\bfseries\centering}{\thesection.}{0.5em}{}
\begin{document}
%
	
\title{Emergence of Nodal-Knot Transitions by Disorder}
	
\author{Ming Gong}
\email{These authors contributed equally to this work}
\affiliation{International Center for Quantum Materials, School of Physics, Peking University, Beijing 100871, China}
\author{Peng-Lu Zhao}
\email{These authors contributed equally to this work}
\affiliation{Department of Physics, University of Science and Technology of China, Hefei, 230026, China}
\author{Hai-Zhou Lu}
\email{Corresponding author: luhz@sustech.edu.cn}
\affiliation{State Key Laboratory of Quantum Functional Materials, Department of Physics, and Guangdong Basic Research Center of Excellence for Quantum Science, Southern University of Science and Technology (SUSTech), Shenzhen 518055, China}
\affiliation{Quantum Science Center of Guangdong-Hong Kong-Macao Greater Bay Area (Guangdong), Shenzhen 518045, China}
\author{Qian Niu}
\affiliation{Department of Physics, University of Science and Technology of China, Hefei, 230026, China}
\affiliation{CAS Key Laboratory of Strongly-Coupled Quantum Matter Physics, University of Science and Technology of China, Hefei, 230026, China}
\affiliation{International Center for Quantum Design of Functional Materials, University of Science and Technology of China, Hefei, 230026, China}
\author{X. C. Xie}
\affiliation{International Center for Quantum Materials, School of Physics, Peking University, Beijing 100871, China}
\affiliation{Interdisciplinary Center for Theoretical Physics and Information Sciences, Fudan University, Shanghai 200433, China}
\affiliation{Hefei National Laboratory, Hefei 230088, China}

\begin{abstract} 
Under certain symmetries, degenerate points in three-dimensional metals form one-dimensional nodal lines. These nodal lines sometimes exhibit intricate knotted structures and have been studied in various contexts. As one of the most common physical perturbations, disorder effects often trigger novel quantum phase transitions. For nodal-knot phases, whether disorder can drive knot transitions remains an open and intriguing question. Employing renormalization-group calculations, we demonstrate that nodal-knot transitions emerge in the presence of weak disorder. Specifically, both chemical-potential-type and magnetic-type disorders can induce knot transitions, resulting in the emergence of distinct knot topologies. The transition can be quantitatively characterized by changes in topological invariants such as the knot Wilson loop integrals. Our findings open up a new avenue for manipulating the topology of nodal-knot phases through disorder effects.
 
\end{abstract} 

\maketitle

\noindent\textbf{1. Introduction}\\
In the 1860s, Lord Kelvin's vortex atom theory introduced the concept of knots to physics, which was further developed by Peter Tait \cite{baez_gauge_1994,simon_topological_2023}. Despite its unfortunate failure, the efforts made by Kelvin and Tait significantly advanced the mathematical study of knot theory \cite{baez_gauge_1994,simon_topological_2023}. A century later, the development of topological quantum field theory revitalized knot theory, paving the way for fault-tolerant quantum computation \cite{atiyah_topological_1988,witten_topological_1988,witten_quantum_1989,blanchet_topological_1995,nayak_non-abelian_2008}. To this day, the development of knot theory has far exceeded expectations, and knot-like states may even play an important role in the formation of the early universe \cite{eto_tying_2024}. 

As the condensed matter counterpart of topological quantum field theories, topological bands serve as ideal platforms for studying the emergence of novel phenomena. Specifically, degenerate points of three-dimensional,(3D) topological bands can form one-dimensional,(1D) nodal lines under the protection of symmetries \cite{burkov_topological_2011,fang_topological_2016}. These nodal lines provide a rich landscape of knots,(nodal-knots) \cite{chang_topological_2017,bi_nodal-knot_2017,ezawa_topological_2017,yan_nodal-link_2017,lian_chern-simons_2017,ahn_band_2018,wu_non-abelian_2019,yang_jones_2020,bouhon_geometric_2020,bouhon_topological_2021-1,bouhon_non-abelian_2020,peng_phonons_2022,jankowski_disorder-induced_2024,chowdhury_phase_2024,kim_unconventional_2024}, and have been studied across a wide range of physical systems\,\cite{bzdusek_nodal-chain_2016,chang_topological_2017,sun_double_2017,li_rules_2018,belopolski_observation_2022,ding_disorder-induced_2024}, including ultra-cold atoms \cite{wang_scheme_2017,deng_probe_2018,tarnowski_measuring_2019,unal_hopf_2019,song_observation_2019,unal_topological_2020}, photonics \cite{leach_knotted_2004,irvine_linked_2008,kedia_tying_2013,yang_observation_2020,jayaseelan_topological_2024,bode_complex_2024}, acoustics \cite{yan_experimental_2018,jiang_experimental_2021,qiu_minimal_2023,zhang_observation_2023,jiang_observation_2024}, and topological circuits \cite{lee_imaging_2020,zhang_tidal_2021,wang_realization_2023}. The focus has also been extended from Hermitian physics to non-Hermitian physics \cite{carlstrom_knotted_2019,yang_non-hermitian_2019,hu_knots_2021,bergholtz_exceptional_2021,patil_measuring_2022,guo_exceptional_2023,zhang_observation_2023,chen_machine_2024,zhu_versatile_2024}.

Regardless of the systems that realize nodal-knots, a fundamental question is how these knotted structures can be tuned through perturbations. Unlike traditional symmetry-protected topological phases \cite{kane_quantum_2005,fu_topological_2007,fu_topological_2007-1}, the flexibility in the ways that 1D nodal lines are embedded into the 3D Brillouin zone to form knots enables the manipulation of nodal-knot structures. On the other hand, it is well-known that disorder effects can significantly renormalize dispersions and even cause localization transitions \cite{lee_disordered_1985,altland_condensed_2010}. Consequently, quantum phase transitions can be triggered, and novel orders may emerge. Therefore, a fundamental and intriguing question naturally arises: Can nodal-knot phases of matter be manipulated by disorder?
\begin{figure}[b]
	\centering
	\includegraphics[width=0.48\textwidth]{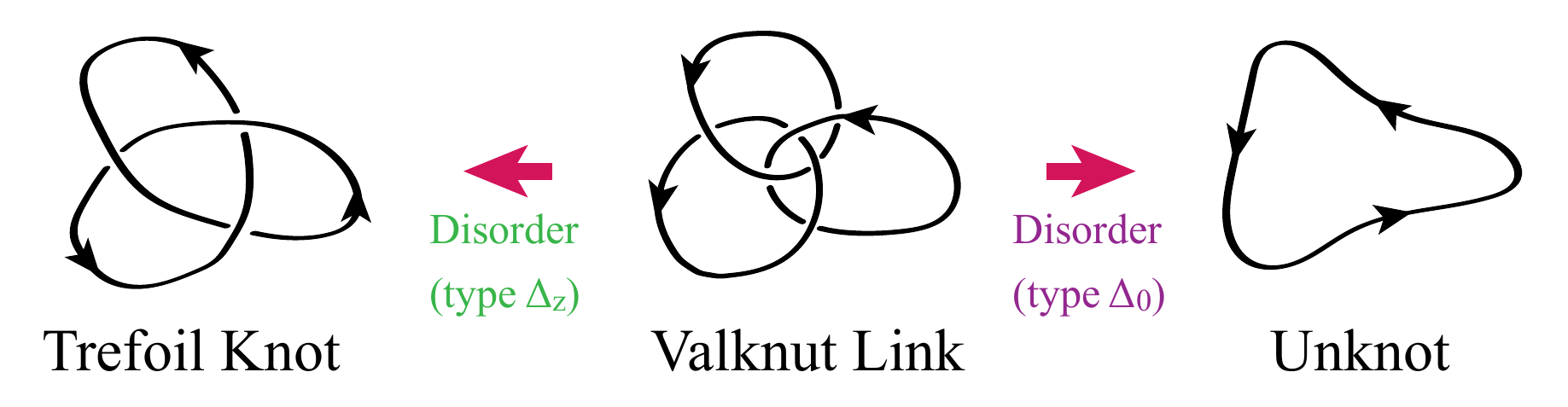}
	\caption{Schematics of the knot transition of nodal-knots in the presence of disorder. The nodal line features a valknut linked structure\,(middle) in the clean limit. Under different types of disorders\,(highlighted with different colors), it can undergo transitions into a trefoil knot\,(left) or a single unknot\,(right).}
	\label{fig1}
\end{figure}

In this work, we address this question by asserting that knot transitions of nodal-knot phases can be triggered by different types of disorder. Employing two-band models of semimetallic systems that host nodal-knot degeneracies, we study the evolution of nodal-knots with various knot topologies under weak disorder. With the help of renormalization-group,(RG) calculations, we find that both chemical-potential-type disorder and magnetic-type disorder can trigger transitions in knot structures. Different types of disorder lead to the evolution of nodal-knots into different knot structures, as sketched in Fig. \ref{fig1}.

Mathematically, these knot transitions can be characterized by the knot Wilson loop integrals, which are knot invariants, and their change can be detected through the de Haas-van Alphen experiment. Our results uncover the fascinating interplay between knots, topological phases, and disorder effects and provide valuable insights for the study of nodal-knots in disordered systems across diverse backgrounds. 
\\
\\
\noindent\textbf{2. Model for 3D nodal-knots}\\
To begin with, we construct the Hamiltonian for 3D nodal-knots. A general two-band Hamiltonian reads $H(\bm k)$\,$=$\,$f_0(\bm k)\sigma_0$\,$+$\,$\sum_{i=x,y,z}f_i(\bm k) \sigma_i$, where $\sigma_i$ are the Pauli matrices. In the presence of chiral symmetry $\mathcal S$\,=\,$\sigma_z$, the Hamiltonian satisfies $\mathcal S H(\bm k) \mathcal S^{-1}$\,=\,$-H(\bm k)$, which forces the $\sigma_0$ and $\sigma_z$ terms to vanish. Then
\begin{equation}
	H(\bm k)=f_x(\bm k)\sigma_x+f_y(\bm k)\sigma_y.
	\label{ham1}
\end{equation}  
The models that host different types of nodal-knots can be obtained by setting \cite{bi_nodal-knot_2017,ezawa_topological_2017}
\begin{align}
f_x(\bm k)={\rm Re}\mathcal{F}_{pq}(\bm k),\,\,\,\, 	f_y(\bm k)={\rm Im}\mathcal{F}_{pq}(\bm k)
\end{align}
where
\begin{align} 
	\nonumber
\mathcal{F}_{pq}(\bm k)=&v_{p}(k_{x}+ik_{y})^{p}+v_{q}[k_z+i(\lambda_1-\lambda_2 k^2)]^q\\
&+m_1+im_2.
\label{Fpq}
\end{align}
Here, $k^2$\,=\,$\sum_{i=x,y,z}k_{i}^2$, $p$ and $q$ are integers. The parameters $v_p$, $v_q$, $\lambda_1$, $\lambda_2$, $m_1$, and $m_2$ govern the shape of the nodal-knot. The degenerate points with $E(\bm k)$\,=\,$\sqrt{f_x^2(\bm k)+f_y^2(\bm k)}$\,=\,$0$ determine the nodal lines in momentum space. When $v_p$\,=\,$v_q$\,$\neq$\,$0$, $m_1$\,=\,$m_2$\,=\,$0$, $\lambda_2$\,=\,$0.5$, and $\lambda_1$\,$>$\,$0$, the nodal line of $\mathcal F_{pq}(k)$ forms the torus knot (or link) of type $(p,q)$, thus different choices of $(p,q)$ in $H(\bm k)$ yield different nodal-knots. For example, $(2,2)$-the Hopf link, $(2,3)$ and $(3,2)$-the trefoil knot, and $(3,3)$-the valknut. When $p$ and $q$ are relatively prime to each other, the nodal-knot is connected (such as the trefoil knot), otherwise, it is disconnected (such as the Hopf link). The nodal lines of $H(\bm k)$ are oriented, similar to the superconducting or superfluid vortex lines \cite{volovik_universe_2003}. The orientation is determined by the unit tangent vector $\bm T(\bm k_0)$ \cite{yang_jones_2020} at point $\bm k_0$. 
\\
\\
\noindent\textbf{3. Disorder and Renormalization}\\
We model the disorder potential by a coupling term $H_{\rm dis}$\,=\,$\int {\rm d}^3{\bm r}U_{\nu}(\bm r)\psi^{\dagger}(\bm r)\sigma_{\nu}\psi(\bm r)$, where ${\nu}$\,=\,$0$ denotes the chemical-potential type disorder and $\sigma_{\nu}$ with ${\nu}$\,=\,$x,y,z$  denote the magnetic-type disorders in three directions \cite{ludwig_integer_1994}. $U_{\nu}(\bm r)$ is the disorder potential of a Gaussian white noise with probability distribution $p[U_{\nu}(\bm r)]$\,$\propto$\,${\rm exp}[{-1/2\Delta_{\nu} U_{\nu}^2(\bm r)}]$ and spatial correlation $\langle U_{\nu}(\bm r)\rangle=0$, $\langle U_{\nu}(\bm r)U_{\rho}(\bm r^{\prime})\rangle$\,=\,$\delta_{\nu\rho}\delta(\bm r-\bm r ^{\prime})$. Through the replica method \cite{wegner_mobility_1979,lee_disordered_1985,altland_condensed_2010}, the fermionic mode $\psi(\bm r,\tau)$ is replicated into $N$ copies as $\psi_{i}(\bm r,\tau)$\,$(i=1\cdots N)$\,(see Supplementary Materials, Sec. I). In this way, the ensemble average results in an effective attractive interaction between replica fields with disorder strength $\Delta_\nu$.  The total action (see Supplementary Materials, Sec. I) comprises a free and a disorder part. We note that although chiral symmetry is externally broken by $\sigma_z$ disorder, it is restored after disorder averaging, since
$\langle U_{z}(\bm r)\rangle=0$.
\begin{figure*}[t]
	\includegraphics[width=0.99\textwidth]{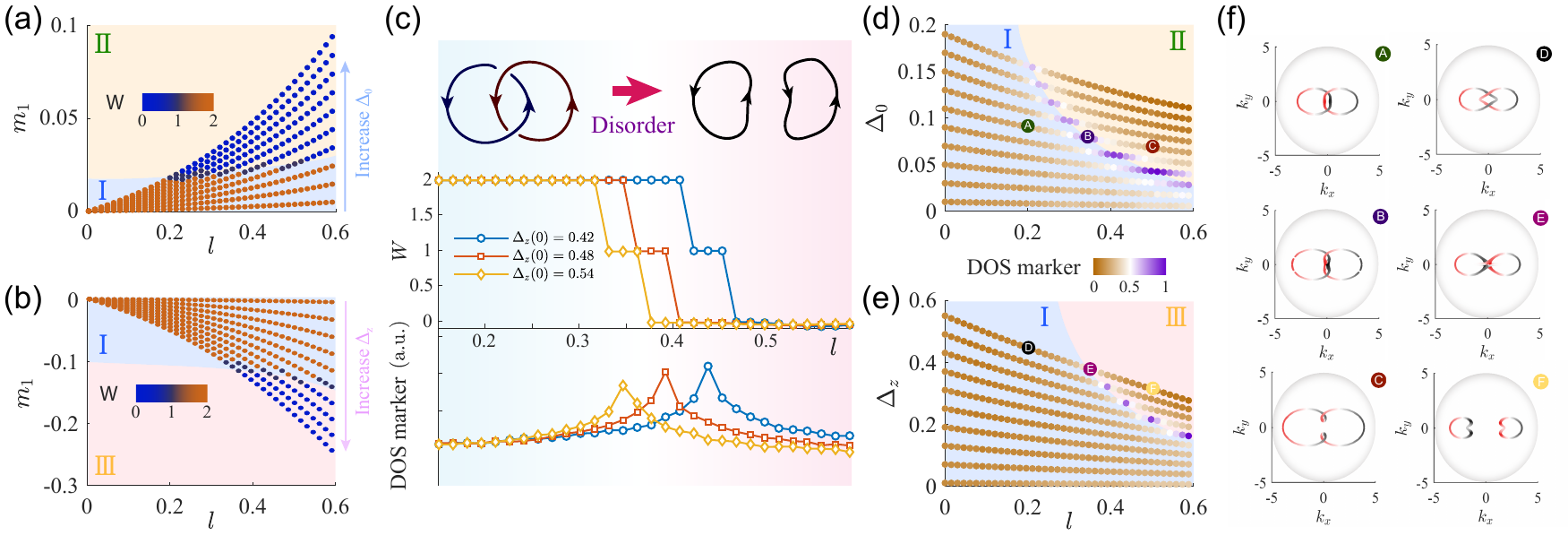}
	\caption{Renormalization-group\,(RG) flows of Hopf linked nodal-knots. (a) and (b), The RG flows of $m_1$ under the renormalization of $\Delta_0$ and $\Delta_z$, respectively. The flow curves are colored by the knot Wilson loop $W$, and colored regions denote I the Hopf link, II the unknot, and III the unlink. (c), The RG flows of the knot Wilson loop\,(upper panel) and the density of states (DOS) marker\,(lower panel) under the renormalization of $\Delta_z$ with different initial values. (d) and (e), The RG flows of $\Delta_0$ and $\Delta_z$, respectively. The flow curves are colored by the DOS marker. The colored regions I, II, and III are the same as that in (a) and (b). The corresponding nodal-knot configurations of representative points labeled A-F on the RG curves are plotted in (f), colored by the $y$-component of unit tangent vector $T_y(\bm k)$.}
	\label{fig2}
\end{figure*}

To study how the band parameters $v_p$, $v_q$, $\lambda_1$, $\lambda_2$, $m_1$, and $m_2$ are renormalized by disorder strength $\Delta_{0,x,y,z}$,  
we then perform the Wilsonian momentum-shell RG calculations for the total action, with details shown in Supplementary Materials \cite{cardy_scaling_1996,kardar_statistical_2007,altland_condensed_2010}. We set the momentum cut off as $\Lambda$. By integrating the fermionic modes inside the momentum-shell defined by $e^{-{\rm d}l}\Lambda$\,$<$\,$|\bm k|$\,$<$\,$\Lambda$ and rescaling the momentum as $\bm k$\,$\to$\,$ e^{-{\rm d}l} \bm k$, we investigate the running of band parameters and disorder strengths $\Delta_{0,x,y,z}$ with respect to the running scale $l$. 
We consider one type of disorder at a time, taking the chemical-potential-type disorder $\Delta_0$ and the magnetic-type disorder $\Delta_z$ as examples. 
Parameters $v_p$, $v_q$, $\lambda_1$, $\lambda_2$ are not renormalized by disorder and only do scale transformations.
The RG equations for $m_1$, and $m_2$ read
\begin{align}
	\nonumber
	{\rm d}m_1/{\rm d}l=&m_1+(-\Delta_0+\Delta_z)\mathcal G_1^x,\quad \\ {\rm d}m_2/{\rm d}l=&m_2+(-\Delta_0+\Delta_z)\mathcal G_1^y ,
	\label{rgeq1}
\end{align}
and the RG equations for $\Delta_0$ and for $\Delta_z$ are
\begin{align}
\nonumber
{\rm d}\Delta_0/{\rm d}l=&-\Delta_0+2\Delta_0^2(\mathcal G^{xx}_2+\mathcal G^{yy}_2),\\
{\rm d}\Delta_z/{\rm d}l=&-\Delta_z-2\Delta_z^2(\mathcal G^{xx}_2+\mathcal G^{yy}_2).
\label{rgeqDz}
\end{align} 
The details for $\mathcal G_1^{(x/y)}$ and $\mathcal G_2^{(xx/yy)}$ can be found in Supplementary Materials

\begin{figure*}[t]
	\includegraphics[width=0.99\textwidth]{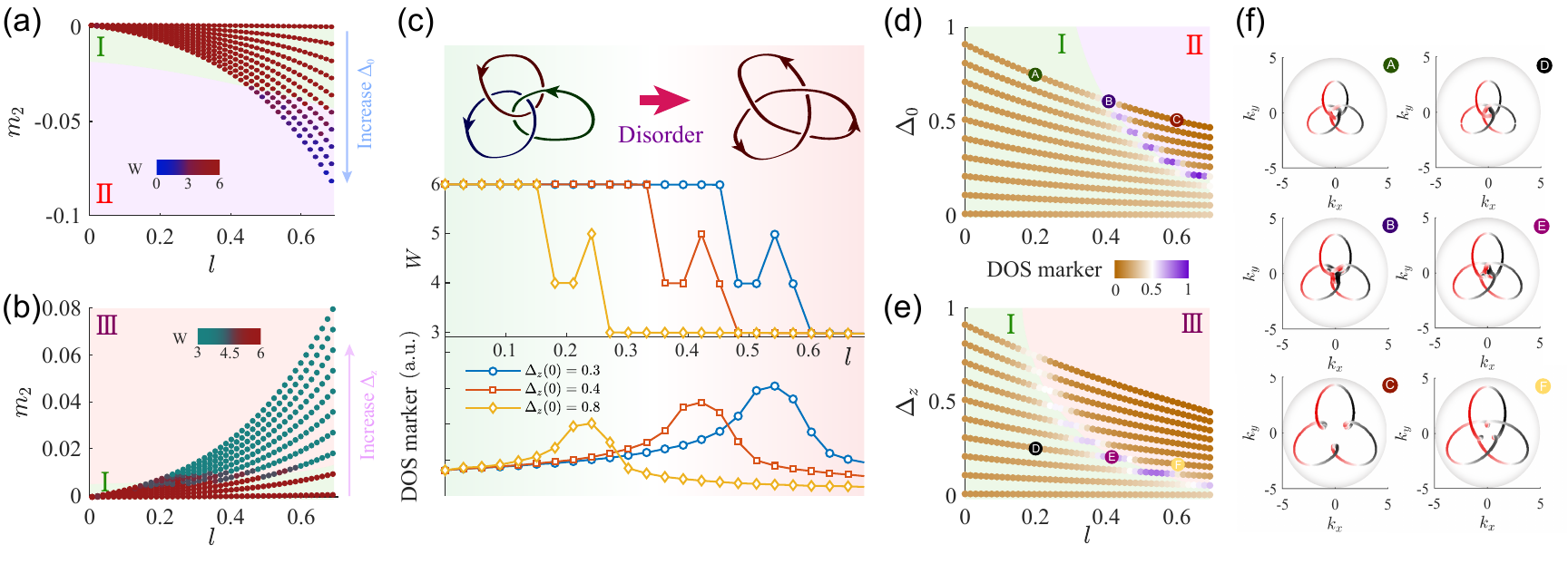}
	\caption{Renormalization-group\,(RG) flows of valknut linked nodal-knots. (a) and (b), The RG flows of $m_2$ under the renormalization of $\Delta_0$ and $\Delta_z$, respectively. The flow curves are colored by the knot Wilson loop $W$, and colored regions denote I the valknut link, II the unknot, and III the trefoil knot. (c), The RG flows of the knot Wilson loop\,(upper panel) and the density of states (DOS) marker\,(lower panel) under the renormalization of $\Delta_z$ with different initial values. (d) and (e), The RG flows of $\Delta_0$ and $\Delta_z$, respectively. The flow curves are colored by the DOS marker. The colored regions I, II, and III are the same as that in (a) and (b). The corresponding nodal-knot configurations of representative points labeled A-F on the RG curves are plotted in (f), colored by the $y$-component of unit tangent vector $T_y(\bm k)$.}
	\label{fig3}
\end{figure*}

The band parameters $v_p$, $v_q$, $\lambda_1$, $\lambda_2$, $m_1$, and $m_2$ renormalized by disorder deviate from their bare values, potentially resulting in distinct knot structures. In other words, disorder may trigger the emergence of knot transitions. We analyse the renormalization of the parameters by numerically solving the RG equations. As an example, we take $p$\,$=$\,$q$\,$=$\,$2$, and set the initial conditions at $l$\,=\,$0$ as $v_p(0)$\,=\,$v_q(0)$\,=\,$0.1$, $m_1(0)$\,=\,$m_2(0)$\,=\,$0$, $\lambda_1(0)$\,=\,$0.4$, and $\lambda_2(0)$\,=\,$0.5$. This set of parameters describe the hopf linked nodal-knot in the clean limit (corresponding to the nodal-knot configuration in the absence of the disorder effect). The momentum cut off takes $\Lambda$\,=\,$5$.
We show the results of the RG flow of $m_1$ in Fig.\,\ref{fig2} a and b under different initial disorder strength $\Delta_0(l=0)$ or $\Delta_z(l=0)$, together with the RG flows of $\Delta_0$ and $\Delta_z$ in  Fig.\,\ref{fig2} d and e. 	Here, $m_2$ keeps zero for that the momentum-shell integration forces $\mathcal G^{y}_1$ to vanish. Clearly, both $\Delta_0$ and $\Delta_z$ can drive the emergence of $m_1$, but in opposite signs. Similar results also hold for $p$\,$=$\,$q$\,$=$\,$3$, the valknut link [Fig.\,\ref{fig3} a and b]. This strongly implies that different types of disorders can trigger different knot transitions. However, only the RG flow of these parameters is not adequate for displaying the features and types of the knot transitions, a more intuitive mathematical object, e.g., a topological invariant, is needed. 
\\
\\
\noindent\textbf{4. Knot invariants and knot transitions in (2,2) and (3,3) systems}\\
Unlike band topological numbers such as TKNN numbers\,\cite{thouless_quantized_1982} and $Z_2$ numbers \,\cite{kane_quantum_2005,fu_topological_2007,fu_topological_2007-1} that unambiguously characterize the symmetry-protected topological phases, the complexity of knots makes it mathematically impossible to find a knot invariant that can provide a one-to-one characterization of the knot configuration. Nevertheless, knot invariants, the jumps of which can adequately reveal nodal-knot transitions, can still be calculated. Here, we adopt the knot Wilson loop integral $W$ to determine the nodal-knot transitions \cite{witten_quantum_1989,lian_chern-simons_2017}. The knot Wilson loop integral is
\begin{eqnarray}
W(L_1,\cdots,L_N)=\frac{1}{\pi}\oint_{\bm{l}\in L_1,\cdots,L_N}\bm{A}(\bm{k})\cdot {\rm d}\bm{l}.
\end{eqnarray}
Here, $L_1,\cdots,L_N$ denotes the nodal loops oriented by $N$, and the orientation of the integral path $\bm l$ is determined by the tangent vector $\bm{T}(\bm k_0)$. $\bm{A}(\bm{k})=-i\langle u_{\bm k}|\partial_{\bm k}|u_{\bm k}\rangle$ is the Berry connection. 
Mathematically, $W(L_1,...,L_N)$ is the sum of the linking numbers $\Psi_{ij}$ of loops $L_i$ and $L_j$ if $i\neq j$, and $\Psi_{ii}$ would be the Gauss linking number of the loop $L_i$ and its frame \cite{baez_gauge_1994,simon_topological_2023}. To faithfully reflect the knot topology, a proper frame should be chosen such that an unknotted loop has no contribution to $W$. Physically, the knot Wilson loop integral $W$ reflects the Berry phase accumulated by electrons as they move along the intertwined nodal lines. For the Hopf-linked nodal-knot, each of the electrons moving along the nodal loops $L_1$ and $L_2$ gains $\pi$ Berry phase, thus $W$$=$$2$.  For the valknut link, each nodal loop interlocks with the other two, and each gains $2\pi$ Berry phase, leading to $W$$=$$6$ in total. 
\begin{figure}[t]
	\includegraphics[width=0.49\textwidth]{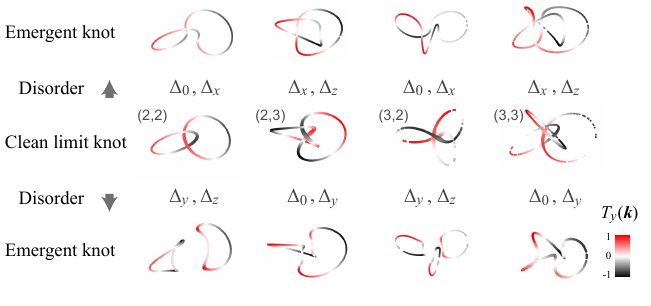}
	\caption{Summary of the emergent knot configurations under different types of disorders. The middle line of the figure shows the clean limit nodal-knots, with various $(p,q)$. Specifically, $(2,2)$-the Hopf link, $(2,3)$ and $(3,2)$-the trefoil knot, and $(3,3)$-the valknut. The $y$-component of unit tangent vector $T_y(\bm k)$ is colored on the nodal-knots.}
	\label{fig4}
\end{figure}
\begin{figure*}[t]
	\centering
	\includegraphics[width=0.9\textwidth]{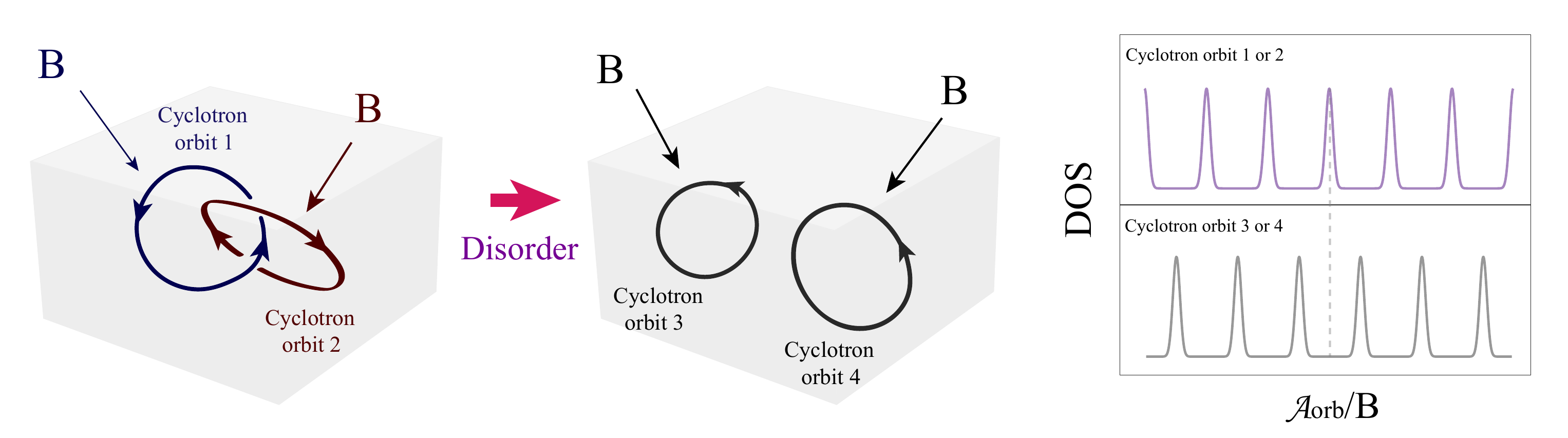}
	\caption{Schematic diagram of utilizing the de Haas-van Alphen oscillation experiment to reflect the nodal-knot transitions. Before the transition, a magnetic field is applied perpendicular to one of the nodal loops (which is also the 1D Fermi surface). Electrons undergo cyclotron motion in momentum space, moving perpendicular to the magnetic field and around the nodal loop (left). The area enclosed by its cyclotron orbit in momentum space is denoted by $A_{\rm orb}$. Sweeping $B$ leads to the oscillations of DOS with period $A_{\rm orb}/B$ (right). Under the renormalization of disorder, the linked structure evolve to be unlinked (middle). The phase of the oscillation with $A_{\rm orb}/B$ shifts by a $\pi$ Berry phase.}
	\label{figRsdH}
\end{figure*}
With this mathematical tool in hand, we can view the nodal-knot transitions more intuitively. We color the RG flows in Fig.\,\ref{fig2} a and b with the calculated knot Wilson loop integral $W$. Moreover, to demonstrate the phase boundaries of nodal-knot transitions more comprehensively, we have supplemented the calculation with the density of states (DOS) marker colored in the RG flow curves of $\Delta_0$ and $\Delta_z$ in Fig.\,\ref{fig2} d and e, which counts the number of states within the energy window $[0,\epsilon]$ with $\epsilon$ an extremely small energy scale. It is worth noting that this DOS marker is different from the real DOS like that was studied in Weyl semimetals \cite{slager_dissolution_2017,pixley_rare-region-induced_2016,roy_global_2018}. There, disorder can broaden the DOS and lead to Fermi surface instabilities, thus fall into the Landau paradigm and different from our study on the Lifshitz transition of the nodal-knots. At the transition points, the nodal lines should touch and then separate and reconnect, leading to the divergence of the DOS markers. In this way, the transition boundaries can be determined jointly by $W$ and the DOS markers. Specifically, we take three curves in Fig.\,\ref{fig2} b and plot them in terms of $W$ and the DOS marker in Fig.\,\ref{fig2} c. It is clear that the knot Wilson loop integral $W$ drops from 2 to 0, and for each curve, the DOS markers show clear peaks at the transition points.

To visualize the renormalized nodal-knot configurations, we select six points on the RG flow curves marked A-F in Fig.\,\ref{fig2} d and e. The corresponding knot configurations are displayed in Fig.\,\ref{fig2} f. Interestingly, although both $\Delta_0$ and $\Delta_z$ drive the transition of $W$ from 2 to 0, the results of these transitions are different. As shown in figure Fig.\,\ref{fig2} f, $\Delta_0$ drives the transition from a Hopf link (A or D) to an unknot (C), while $\Delta_z$ drives the transition to an unlink (F). This difference stems from their renormalization effect to the parameters $m_1$ and $m_2$ with opposite sign. Therefore, we can divide the RG flow diagrams into regions I (Hopf link), II (unknot), and III (unlink) [shown in Fig.\,\ref{fig2} a,b and d,e].
 
Similar to the Hopf link, the valknut nodal-knot with $p$\,$=$\,$q$\,$=$\,$3$ also experience different transitions under different types of disorder. In parallel to the Hopf link, we show the results of RG flow and the corresponding $W$ and DOS markers [Fig.\,\ref{fig3} a-e]. Different from the Hopf link, only $m_2$ survives. We see that under the renormalization of $\Delta_0$, $W$ jumps from 6 to 0, while under $\Delta_z$, $W$ jumps from 6 to 3, which is distinct from the Hopf links. Moreover, representative nodal-knot configurations shown in Fig.\,\ref{fig3} f suggests that we can similarly divide the RG flow into regions 
\rm{I}, \rm{II}, and \rm{III}, which describe the nodal-knot configurations of valknut, unknot, and trefoil knot, respectively. Note that under $\Delta_z$, the originally linked valknut evolve into the trefoil knot, which is a knotted structure, in sharp contrast to the case of the Hopf link which evolves into either unknot or unlink. 
\begin{table}[h]
\centering
\resizebox{\linewidth}{!}{
\begin{tabular}{lcccc}
\toprule
\multicolumn{1}{c}{\multirow{2}{*}{Disorder type}} & \multicolumn{4}{c}{$(p,q)$} \\ \cmidrule(lr){2-5} 
			& $(2,2)$ &  $(2,3)$ & $(3,2)$ & $(3,3)$ \\
	\midrule
	Clean & Hopf link & Trefoil knot & Trefoil knot & Valknut \\
	$\Delta_0$ & Unknot & Unknot & Unknot & Unknot \\
	$\Delta_x$ & Unknot & Hopf link & Unknot & Trefoil knot \\
	$\Delta_y$ & 2-Unlink & Unknot & 3-Unlink & Unknot \\
	$\Delta_z$ & 2-Unlink & Hopf link & 3-Unlink & Trefoil knot \\
	\bottomrule
	\end{tabular}
}
\caption{Summary of the emergent knot transitions for various $(p,q)$ triggered by different types of disorders.}
\label{tab-label-1}
\end{table}

As summarized in Fig.\,\ref{fig4} and Table. \ref{tab-label-1}, we can conclude that the ability of different types of disorder to induce different types of nodal-knot transitions is universal. Especially, the presence of magnetic disorders can diversify the knot configurations (see the table in Table. \ref{tab-label-1}), giving rise to the emergence of non-trivial nodal-knots from their clean limit.  This finding plays a crucial role in realizing and manipulating topological knotted phases in various systems in the future. 
\\
\\
\noindent\textbf{5. Phase shifts in quantum oscillation experiments}\\	
Experimentally, determining the transition of the Fermi surface topologies (i.e., Lifshitz transition) is typically accomplished by combining transport measurements with quantum oscillation experiments. Therefore, we propose a potentially feasible way to reflect the nodal-knot structures and their transitions through de Haas-van Alphen oscillation or Shubnikov-de Haas oscillation experiment \cite{hu_evidence_2016,ali_butterfly_2016,pezzini_unconventional_2018,shi_quantum_2024}, as demonstrated in Fig. \ref{figRsdH}. Taking the Hopf link as an example, before the transition occurs, a magnetic field is applied perpendicular to one of the nodal loops, which is the 1D Fermi surface of the system at $\mu=0$. According to the semiclassical equations of motion \cite{xiao_berry_2010}
\begin{align}
\nonumber
\dot{\bm{r}}=&\partial \epsilon (\bm k)/\hbar\partial \bm k-\dot{\bm{k}}\times\bm \Omega(\bm k),\\
\dot{\bm{k}}=&-e/\hbar \bm E-e/\hbar \dot{\bm{r}}\times\bm B,
\end{align}
electrons undergo cyclotron motion in momentum space, moving perpendicular to the magnetic field and around the nodal loop, as indicated by the loops of different colors in the left panel of Fig.\,\ref{figRsdH}. The area enclosed by its cyclotron orbit in momentum space is denoted by $A_{\rm orb}$. When the magnetic field changes, the DOS exhibits periodic oscillations with a period $A_{\rm orb}/B$, as shown in the right panel of Fig.\,\ref{figRsdH}. Under the renormalization of disorder, the linked structure evolves to be unlinked, as shown in the middle panel of Fig.\,\ref{figRsdH}. Applying a magnetic field in this situation still results in quantum oscillations, but the phase of the oscillation with $A_{\rm orb}/B$ may shift. When two nodal loops are linked together, the cyclotron motion orbit of electrons on one nodal loop will encircle the other nodal loop. According to the Landau quantization rule
\begin{eqnarray}
A_{\rm orb}(\bm k_{\bm B})=2\pi eB(n+1/2+\phi_{B}),
\end{eqnarray}
where $A_{\rm orb}(\bm k_{\bm B})$ is the area of the cyclotron orbit with $\bm k_{B}$ the wave vector along the $\bm B$ direction, and $\phi_{B}$ is the Berry phase accumulated by the electrons during the cyclotron motion \cite{mikitik_manifestation_1999,xiao_berry_2010}. From the previous analysis, the knot Wilson loop integral precisely corresponds to the sum of the Berry phases $\phi_B$ accumulated by each cyclotron orbit. Therefore, we see that the Berry phase $\phi_B$
is $\pi$ in this case. Under the renormalization of disorder, the linked nodal loops are separated, resulting in $\phi_B=0$. The difference in the Berry phase $\phi_B$ thus induces the $\pi$ Berry phase shift in the de Haas-van Alphen oscillation experiment as depicted in the right panel of Fig. \ref{figRsdH}. Therefore, these accumulated Berry phases $\phi_B$ for each nodal loop are unique for knotted nodal lines. If all the information of $\phi_B$ can be obtained, the nodal-knot structure can be determined to a large extent.
\\
\\
\noindent\textbf{6. Discussions}\\
Nodal line semimetals with non-trivial knotted or linked structures have been confirmed to exist in solid materials, such as Co$_2$MnGa through ARPES experiments \cite{chang_topological_2017,belopolski_observation_2022}. In the presence of disorder, the ARPES observations would be affected by the renormalization of the low-energy band structure. Our work predicts the possibility of disorder induced nodal-knot transitions under doping with magnetic or non-magnetic disorders, which awaits further experimental verification. 

Aside from the solid materials, it is also achievable to introduce and control disorders in artificial systems such as cold atoms or acoustic and optical systems \cite{deng_probe_2018,yang_observation_2020,zhang_observation_2023}. Especially, introducing noise in cold atoms, or utilizing the optical speckle field can simulate the disorder effect \cite{kondov_three-dimensional_2011}. Moreover, optical Raman lattice can simulate the spin-orbital coupling, which can achieve the magnetic disorder potentials \cite{song_observation_2019}. In these ways, magnetization directions of the disorders can be controlled more easily. Therefore, our work provides theoretical support for future studies of knotted topological phases and disorder-induced knot transitions in these systems.

\quad
\\
\noindent
\textbf{Conflict of interest} \\
The authors declare no competing interest.
\quad
\\

\noindent\textbf{Acknowledgments}\\
This work was supported by the National Key R$\&$D Program of China (2022YFA1403700), the Innovation Program for Quantum
Science and Technology (2021ZD0302400), the National Natural
Science Foundation of China (12350402, 12304074, 12234017,
and 12525401), the Guangdong Province (2020KCXTD001), the
Guangdong Basic and Applied Basic Research Foundation (2023B0303000011), the Guangdong Provincial Quantum Science
Strategic Initiative (GDZX2201001 and GDZX2401001), the
Science, Technology and Innovation Commission of Shenzhen
Municipality (ZDSYS20190902092905285), and the New Cornerstone Science Foundation through the XPLORER PRIZE, and Center for Computational Science and Engineering of SUSTech. Ming Gong was also supported by the China National Postdoctoral Program for Innovative Talents (BX20240004). We acknowledge helpful discussions with Robert-Jan Slager, Qing-Feng Sun, Hua Jiang, Chui-Zhen Chen, Zhenyu Xiao, and Xinchi Zhou.
\quad
\\
\\
\noindent\textbf{Author Contributions}\\
Ming Gong and Peng-Lu Zhao conceived the idea from discussions with Hai-Zhou Lu, Ming Gong performed calculations with assistance from Peng-Lu Zhao, Ming Gong and Peng-Lu Zhao wrote the manuscript with contributions from all authors. Hai-Zhou Lu, Qian Niu, and X. C. Xie supervised the project.
\quad
\\
\\

\end{document}


\title{Supplementary Materials for ``Emergence of Nodal-Knot Transitions by Disorder''}

\author{Ming Gong}
\email{These authors contributed equally to this work}
\affiliation{International Center for Quantum Materials, School of Physics, Peking University, Beijing 100871, China}
\author{Peng-Lu Zhao}
\email{These authors contributed equally to this work}
\affiliation{Department of Physics, University of Science and Technology of China, Hefei, 230026, China}
\author{Hai-Zhou Lu}
\email{Corresponding author: luhz@sustech.edu.cn}
\affiliation{State Key Laboratory of Quantum Functional Materials, Department of Physics, and Guangdong Basic Research Center of Excellence for Quantum Science, Southern University of Science and Technology (SUSTech), Shenzhen 518055, China}
\affiliation{Quantum Science Center of Guangdong-Hong Kong-Macao Greater Bay Area (Guangdong), Shenzhen 518045, China}
\author{Qian Niu}
\affiliation{Department of Physics, University of Science and Technology of China, Hefei, 230026, China}
\affiliation{CAS Key Laboratory of Strongly-Coupled Quantum Matter Physics, University of Science and Technology of China, Hefei, 230026, China}
\affiliation{International Center for Quantum Design of Functional Materials, University of Science and Technology of China, Hefei, 230026, China}
\author{X. C. Xie}
\affiliation{International Center for Quantum Materials, School of Physics, Peking University, Beijing 100871, China}
\affiliation{Interdisciplinary Center for Theoretical Physics and Information Sciences, Fudan University, Shanghai 200433, China}
\affiliation{Hefei National Laboratory, Hefei 230088, China}

	\maketitle
	
	\tableofcontents

\section{Fermionic system under quenched disordered: the replica treatment}

In the presence of a quenched disorder potential $A_mV_m(\bm r)$ (here, $A_m$ may be some matrix to represent different types of disorder), the partition function for a non-interacting fermion system reads \cite{wegner_mobility_1979,lee_disordered_1985,altland_condensed_2010}:
\begin{equation}
\mathcal{Z}=\int \mathcal{D}[\psi,\bar \psi]e^{-S_{0}-S_{\rm dis}}
\end{equation} 
where the free part of the action is:
\begin{equation}
S_{0}=-\int d^3{\bm r}d^3{\bm r^\prime}d\tau d\tau^\prime \bar{\psi}({\bm r},\tau)\mathbf{G}_{0}^{-1}({\bm r},\tau,{\bm r^\prime},\tau^\prime)\psi({\bm r^\prime},\tau^\prime)
\end{equation}
with $\mathbf{G}_{0}$ the Green's function of free fermions, and $\tau=it$. The disorder part:
\begin{equation}
S_{\rm dis}=\int d^3{\bm r}d\tau \bar \psi({\bm r},\tau)\left[\sum\limits_{m=1}^{M} A_mV_m({\bm r}) \right]\psi({\bm r},\tau).
\end{equation}
The total free energy of the system is the disorder averaged logarithmic partition function:
\begin{align}
	\nonumber
\mathcal{W}&=-\langle {\rm ln}\mathcal{Z} \rangle _{\rm dis}\\
			&=\int\mathcal{D}[V_1]\cdots \mathcal{D}[V_M]P[V_1]\cdots P[V_M]{\rm ln} \mathcal{Z}
\end{align}
with $M$ types of disorders, and $P[V_m]$ is the distribution functional of the disorder potential. Here we consider that the disorders have no spatial correlations, i.e., $\langle V_m({\bm r})V_n({\bm r^\prime}) \rangle=\delta_{mn}\delta({\bm r}-{\bm r^\prime})$, and fluctuates with a Gaussian white noise distribution. $P[V_m]$ is assumed to have the form:
\begin{equation}
			p[V_m] \propto e^{-\frac{1}{2\Delta_m}V_m^{2}}.
\end{equation} 
Here $\Delta_m$ represents the fluctuation ``width'' of the disorder potential $A_mV_m(\mathbf r)$, which can be understood as the disorder strength. Using the replica method:
\begin{equation}
\langle {\rm ln}\mathcal{Z} \rangle_{\rm dis}=\mathop{\rm lim} \limits_{N \rightarrow 0}\langle\frac{\mathcal{Z}^{N}-1}{N}\rangle_{\rm dis},
\end{equation}
the original system is replaced by $N$ identical copies that share the same disorder potential. Here we first evaluate $\mathcal{Z}^{N}$
\begin{align}
\langle \mathcal{Z}^{N} \rangle_{\rm dis}=\int \mathcal{D}[V]P[V]\mathcal{D}[\bar \psi_{1}, \psi_{1};\cdots ;\bar \psi_{N}, \psi_{N}]e^{\int \bar{\psi}_{1}\mathbf{G}_{0}^{-1}\psi_{1}+\bar{\psi}_{2}\mathbf{G}_{0}^{-1}\psi_{2}+\cdots+\bar{\psi}_{N}\mathbf{G}_{0}^{-1}\psi_{N}-\sum\limits_{m=1}^{M}(\bar \psi_{1}A_m\psi_{1}+\cdots+\bar \psi_{N}A_m\psi_{N})V_m }.
\end{align}
After averaging over the disorder potentials, the effective partition function becomes:
\begin{align}
\langle \mathcal{Z}^{N} \rangle_{\rm dis}=\int\mathcal{D}[\bar \psi_{1}, \psi_{1};\cdots ;\bar \psi_{N}, \psi_{N}]e^{\int \bar{\psi}_{1}\mathbf{G}_{0}^{-1}\psi_{1}+\bar{\psi}_{2}\mathbf{G}_{0}^{-1}\psi_{2}+\cdots+\bar{\psi}_{N}\mathbf{G}_{0}^{-1}\psi_{N}+\sum\limits_{i,j=1}^{N} \sum\limits_{m=1}^{M} \frac{\Delta_m}{2}\bar{\psi_{i}}A_m\psi_{i}\bar{\psi}_{j}A_m\psi_{j}}.
\end{align}
At this stage, the effective action of the replicated system is:
\begin{align}
\nonumber
S_{\rm eff}=&-\int d^3{\bm r}d^3{\bm r^\prime}d\tau d\tau^\prime\sum\limits_{i=1}^{N} \bar{\psi}_i({\bm r},\tau)\mathbf{G}_{0}^{-1}({\bm r},\tau,{\bm r^\prime},\tau^\prime)\psi_i({\bm r^\prime},\tau^\prime)\\
&-\int d^3{\bm r}d\tau d\tau^\prime\sum\limits_{i,j=1}^{N} \sum\limits_{m=1}^{M}\frac{\Delta_m}{2}(\bar{\psi_{i}}A_m\psi_{i})_{{\bm r},\tau}(\bar{\psi}_{j}A_m\psi_{j})_{{\bm r},\tau^\prime}.
\end{align}
The average over the disorder potential leads to an effective attractive interaction among the replicated fields. At the end of the calculation, the limit $N\to0$ should be taken.
\section{Fourier transformations}
Fourier transforms of field operators in real time:
\begin{align}
\nonumber
&\psi({\bm k},\omega)= \int d^{3}{\bm r}dt e^{-i {\bm k}\cdot{\bm r}+i \omega t}\psi({\bm r}, t)\\
&\psi({\bm r},t)= \int \frac{d^{3}{\bm k}d\omega}{(2\pi)^4}  e^{i {\bm k}\cdot{\bm r}-i \omega t}\psi({\bm k},\omega)
\end{align}
Fourier transforms of field operators in imaginary time ($t\to-i\tau$, $\omega_n=\frac{(2n+1)\pi}{\beta}$ is the Matsubara frequency):
\begin{align}
\nonumber
&\psi({\bm k},\omega_n)=\int_V  \int_{0}^{\beta} d^{3}{\bm r}d\tau e^{-i {\bm k}\cdot{\bm r}+i \omega_n \tau}\psi({\bm r}, \tau)\\
&\psi({\bm r},\tau)=\frac{1}{\beta}\sum_{n}\int_\Omega  \frac{d^{3}{\bm k}}{(2\pi)^3}  e^{i {\bm k}\cdot{\bm r}-i \omega_n \tau}\psi({\bm k},\omega_n).
\end{align}
Suppose that the Hamiltonian of a non-interacting system is $h_0$ in the absence of disorder, and $h_0$ is time-independent and local, which can be denoted as $h_0=h_0({\bm r},\nabla{\bm r},\cdots)$. The effective action in the presence of various types of disorders can be written as
\begin{align}
\nonumber
S_{\rm eff}=&\int d^3{\bm r}d\tau\sum\limits_{i=1}^{N}\bar{\psi}_i({\bm r},\tau)(\partial_\tau +h_0)\psi_i({\bm r},\tau)\\
&-\int d^3{\bm r}d\tau d\tau^\prime\sum\limits_{i,j=1}^{N} \sum\limits_{m=1}^{M}\frac{\Delta_m}{2}(\bar{\psi_{i}}A_m\psi_{i})_{{\bm r},\tau}(\bar{\psi}_{j}A_m\psi_{j})_{{\bm r},\tau^\prime}.
\end{align}
The Fourier transformed $S_{\rm eff}$ (imaginary time) reads
\begin{align}
\nonumber
S_{\rm eff}=&\frac{1}{\beta}\sum_n\int \frac{d^3{\bm k}}{(2\pi)^3}\sum\limits_{i=1}^{N} \bar{\psi}_i({\bm k},\omega_n)[-i\omega_n +h_0({\bm k})]\psi_i({\bm k},\omega_n)\\
&-\frac{1}{\beta^2}\sum_{n,n^\prime}\int \frac{d^3{\bm k_1}}{(2\pi)^3} \frac{d^3{\bm k_2}}{(2\pi)^3} \frac{d^3{\bm k_3}}{(2\pi)^3} \frac{d^3{\bm k_4}}{(2\pi)^3}(2\pi)^3\delta({\bm  k_1+\bm k_3-\bm k_2-\bm k_4})\nonumber\\ 
&\times \sum\limits_{i,j=1}^{N} \sum\limits_{m=1}^{M}\frac{\Delta_m}{2}[\bar{\psi_{i}}(\bm k_1,\omega_n)A_m\psi_{i}(\bm k_2,\omega_n)]\cdot[\bar{\psi}_{j}(\bm k_3,\omega_{n^\prime})A_m\psi_{j}(\bm k_4,\omega_{n^\prime})].
\end{align}

In the following analyses of two-band models, all types of disorders are considered. The disorder potential is  
\begin{equation}
U({\bm r})=U_0({\bm r})\sigma_0+U_x({\bm r})\sigma_x+U_y({\bm r})\sigma_y+U_z({\bm r})\sigma_z.
\end{equation} 
where the first term is the chemical-potential-type disorder and the last three terms are the magnetic-type disorders in the three directions \cite{ludwig_integer_1994}. The distribution is $p[U_m({\bm r})]\propto {\rm exp}[-\frac{1}{2\Delta_m} d^3{\bm r}U_m({\bm r})^2]$ for $m=0,x,y,z$. After performing the replica method, the effective action $S_{\rm eff}=S_0+S_1$ includes both the non-interacting part
\begin{align}
S_{0}=\int d^3{\bm r}d\tau\sum\limits_{i=1}^{N}\bar{\psi}_i({\bm r},\tau)[\partial_\tau +H({\bm r}, {\nabla\bm r},\cdots)-\mu]\psi_i({\bm r},\tau)
\end{align}
and the disorder induced effective interaction part
\begin{align}
S_{1}=-\int d^3{\bm r}d\tau d\tau^\prime\sum\limits_{i,j=1}^{N} \sum\limits_{m=0,x,y,z}\frac{\Delta_m}{2}(\bar{\psi_{i}}\sigma_m\psi_{i})_{{\bm r},\tau}(\bar{\psi}_{j}\sigma_m\psi_{j})_{{\bm r},\tau^\prime}.
\end{align}
\section{RG step 1: coarse grain}
We use the momentum-shell RG method \cite{cardy_scaling_1996,kardar_statistical_2007,altland_condensed_2010,goswami_quantum_2011,luo_quantum_2018,zhao_theory_2021,isobe_supermetal_2019} to analyse the knot transitions of nodal-knot semimetals with Hamiltonian $	H(\bm k)=f_x(\bm k)\sigma_x+f_y(\bm k)\sigma_y$ under the weak disorder regime. The disorder strengths $\Delta_m$ ($m=0,x,y,z$) can be viewed as small parameters.

The fermionic field $\psi({\bm k},\omega_n)$ is decoupled into a ``slow'' mode $\psi_<({\bm k},\omega_n)$ and a ``fast'' mode $\psi_>({\bm k},\omega_n)$, which is defined as
\begin{align}
\nonumber
\psi_<({\bm k},\omega_n)&=\Theta(\frac \Lambda b -|k|)\psi({\bm k},\omega_n),\\
\psi_>({\bm k},\omega_n)&=\Theta(\Lambda-|k|)\Theta(|k|-\frac \Lambda b)\psi({\bm k},\omega_n).
\end{align}
Here, $\Lambda$ is the momentum cutoff. $b=e^l$ is the scaling factor with $l$ the running scale parameter. The ``fast'' mode lies on the momentum-shell that is to be integrated out.
		
The effective action can be rewritten as:
\begin{align}
\nonumber
& S_{0}=S_{0,\textless}+S_{0,\textgreater}\\
& S_{1}=S_{1,\textless}+S_{1,\textgreater},
\end{align}
where
\begin{align}
\nonumber
S_{0,<}&=\int d^3{\bm r}d\tau\sum\limits_{i=1}^{N}{\bar\psi}_{i,<}({\bm r},\tau)(\partial_\tau +H-\mu)\psi_{i,<}({\bm r},\tau)\\ 
&=\frac{1}{\beta}\sum_n\int \frac{d^3{\bm k}}{(2\pi)^3}\sum\limits_{i=1}^{N} \bar{\psi}_{i,<}({\bm k},\omega_n)[-i\omega_n +H({\bm k})-\mu]\psi_{i,<}({\bm k},\omega_n)\\\nonumber
S_{0,>}&=\int d^3{\bm r}d\tau\sum\limits_{i=1}^{N}{\bar\psi}_{i,>}({\bm r},\tau)(\partial_\tau +H-\mu)\psi_{i,>}({\bm r},\tau)\\ 
&=\frac{1}{\beta}\sum_n\int \frac{d^3{\bm k}}{(2\pi)^3}\sum\limits_{i=1}^{N} \bar{\psi}_{i,>}({\bm k},\omega_n)[-i\omega_n +H({\bm k})-\mu]\psi_{i,>}({\bm k},\omega_n).
\end{align}
\begin{align}
\nonumber
S_{1,<}=&-\frac{1}{\beta^2}\sum_{n,n^\prime}\int \frac{d^3{\bm k_1}}{(2\pi)^3} \frac{d^3{\bm k_2}}{(2\pi)^3} \frac{d^3{\bm k_3}}{(2\pi)^3} \frac{d^3{\bm k_4}}{(2\pi)^3}(2\pi)^3\delta({\bm  k_1+\bm k_3-\bm k_2-\bm k_4})\\
&\times \sum\limits_{i,j=1}^{N} \sum\limits_{m=0,x,y,z}\frac{\Delta_m}{2}[{\bar\psi_{i,<}}(\bm k_1,\omega_n)\sigma_m\psi_{i,<}(\bm k_2,\omega_n)]\cdot[{\bar\psi}_{j,<}(\bm k_3,\omega_{n^\prime})\sigma_m\psi_{j,<}(\bm k_4,\omega_{n^\prime})].
\end{align}
and
\begin{align} 
\nonumber
S_{1,>}=&-\frac{1}{\beta^2}\sum_{n,n^\prime}\int \frac{d^3{\bm k_1}}{(2\pi)^3} \frac{d^3{\bm k_2}}{(2\pi)^3} \frac{d^3{\bm k_3}}{(2\pi)^3} \frac{d^3{\bm k_4}}{(2\pi)^3}(2\pi)^3\delta({\bm  k_1+\bm k_3-\bm k_2-\bm k_4})\\
&\times \sum\limits_{i,j=1}^{N} \sum\limits_{m=0,x,y,z}\sum_{\alpha\beta\gamma\delta}^{\text{at least one \textgreater}}\frac{\Delta_m}{2}[{\bar\psi_{i,\alpha}}(\bm k_1,\omega_n)\sigma_m\psi_{i,\beta}(\bm k_2,\omega_n)]\cdot[{\bar\psi}_{j,\gamma}(\bm k_3,\omega_{n^\prime})\sigma_m\psi_{j,\delta}(\bm k_4,\omega_{n^\prime})].
\end{align}
The functional integral can be rewritten explicitly in terms of $>$ and $<$ as
\begin{align}
\nonumber
\langle \mathcal{Z}^{N} \rangle_{\rm dis} =\mathcal{Z}_{\rm eff}&=\int \mathcal{D}[\bar\psi_{\textless},\psi_{\textless};\bar\psi_{\textgreater},\psi_{\textgreater}]e^{-S_{\rm eff}}\\ \nonumber
&=\int \mathcal{D}[\bar\psi_{\textless},\psi_{\textless};\bar\psi_{\textgreater},\psi_{\textgreater}]e^{-S_{0,\textless}-S_{0,\textgreater}-S_{1,\textless}-S_{1,\textgreater}}\\ \nonumber
&=\int \mathcal{D}[\bar\psi_{\textless},\psi_{\textless}]e^{-S_{0,\textless}-S_{1,\textless}} \int \mathcal{D}[\bar\psi_{\textgreater},\psi_{\textgreater}]e^{-S_{0,\textgreater}-S_{1,\textgreater}} \\ \nonumber
&=\mathcal{Z}_{0,\textgreater}\int\mathcal{D}[\bar\psi_{\textless},\psi_{\textless}]e^{-S_{0,\textless}-S_{1,\textless}}\langle e^{-S_{1,\textgreater}}\rangle_{0,\textgreater} \\
&=\mathcal{Z}_{0,\textgreater}\int\mathcal{D}[\bar\psi_{\textless},\psi_{\textless}]e^{-S_{0,\textless}-S_{1,\textless}}e^{-\langle S_{1,\textgreater} \rangle_{0,\textgreater}+\frac 12 (\langle S^{2}_{1,\textgreater} \rangle_{0,\textgreater}-\langle S_{1,\textgreater }\rangle_{0,\textgreater}^{2})+\cdots}.
\end{align}
where the average over the fast modes is defined as
\begin{equation}
\langle \cdots \rangle_{0,\textgreater}=\frac{\int \mathcal{D}[\bar\psi_{\textgreater},\psi_{\textgreater}] e^{-S_{0,\textgreater}} (\cdots)}{\int \mathcal{D}[\bar\psi_{\textgreater},\psi_{\textgreater}] e^{-S_{0,\textgreater}}}.
\end{equation}
and $\mathcal{Z}_{0,\textgreater}=\int \mathcal{D}[\bar\psi_{\textgreater},\psi_{\textgreater}] e^{-S_{0,\textgreater}}$. 
		
\begin{figure}[b]
\centering
\includegraphics[width=0.4\textwidth]{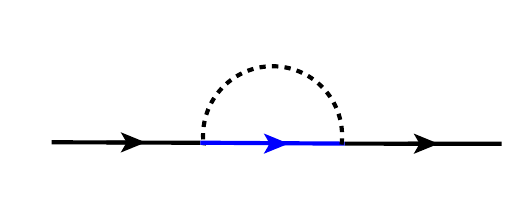}
\caption{Diagram for the first order perturbation. Black legs denote the ``slow'' modes and the blue legs denote the ``fast'' modes that should be integrated.}
\label{fig_fstorder}
\end{figure}
To the first order in $\Delta_m$, only one diagram should be considered as shown in Fig. \ref{fig_fstorder}. The modification of $S_{0,<}$ from $S_{1,>}$ is 
\begin{align}
\delta S_{0,<}=\langle S_{1,>}\rangle_{0,>},
\end{align}
where 
\begin{align}
\nonumber
\langle S_{1,>}\rangle_{0,>}=&-\frac{1}{\beta^2}\sum_{n,n^\prime}\int \frac{d^3{\bm k_1}}{(2\pi)^3} \frac{d^3{\bm k_2}}{(2\pi)^3} \frac{d^3{\bm k_3}}{(2\pi)^3} \frac{d^3{\bm k_4}}{(2\pi)^3}(2\pi)^3\delta({\bm  k_1+\bm k_3-\bm k_2-\bm k_4})\times \\ \nonumber
&2\sum\limits_{i,j=1}^{N} \sum\limits_{m=0,x,y,z}\frac{\Delta_m}{2}{\bar\psi_{i,<}}(\bm k_1,\omega_n)\sigma_m\langle\psi_{i,>}(\bm k_2,\omega_n){\bar\psi}_{j,>}(\bm k_3,\omega_{n^\prime})\rangle_{0,>}\sigma_m\psi_{j,<}(\bm k_4,\omega_{n^\prime})\\ \nonumber
=&\frac{1}{\beta}\sum_{n}\int \frac{d^3{\bm k_1}}{(2\pi)^3} \frac{d^3{\bm k_2}}{(2\pi)^3}\sum\limits_{i=1}^{N} \sum\limits_{m=0,x,y,z}\Delta_m{\bar\psi_{i,<}}(\bm k_1,\omega_n)\sigma_m[G_0(\bm k_2,\omega_n)]_{(\bm k_2\in >)}\sigma_m\psi_{i,<}(\bm k_1,\omega_{n})\\
=&\frac{1}{\beta}\sum_{n}\int_< \frac{d^3{\bm k_1}}{(2\pi)^3} \sum\limits_{i=1}^{N} \sum\limits_{m=0,x,y,z}\Delta_m{\bar\psi_{i,<}}(\bm k_1,\omega_n)\sigma_m\int_>\frac{d^3{\bm k_2}}{(2\pi)^3}\frac{\mathbb I}{i\omega_n+\mu-f_x(\bm k_2)\sigma_x-f_y(\bm k_2)\sigma_y}\sigma_m\psi_{i,<}(\bm k_1,\omega_{n}).
\end{align}
The integration of $\bm k_2$ is restricted on the momentum shell $\Lambda e^{-dl}<|k_2|<\Lambda$. Thus the above equation becomes
\begin{small}
\begin{align}
\nonumber
\delta S_{0,<}&=\frac{1}{\beta}\sum_{n}\int_< \frac{d^3{\bm k_1}}{(2\pi)^3} \sum\limits_{i=1}^{N} \sum\limits_{m=0,x,y,z}\Delta_m{\bar\psi_{i,<}}(\bm k_1,\omega_n)\sigma_m\int\frac{\Lambda^3dl{\rm sin}\theta d\theta d\phi}{(2\pi)^3}\frac{\mathbb I}{i\omega_n+\mu-f_x\sigma_x-f_y\sigma_y}\sigma_m\psi_{i,<}(\bm k_1,\omega_{n})\\
&=\frac{1}{\beta}\sum_{n}\int_< \frac{d^3{\bm k_1}}{(2\pi)^3} \sum\limits_{i=1}^{N} \sum\limits_{m=0,x,y,z}\Delta_m{\bar\psi_{i,<}}(\bm k_1,\omega_n)\int\frac{\Lambda^3dl{\rm sin}\theta d\theta d\phi}{(2\pi)^3}\frac{-i\omega_n-\mu-f_x\sigma_m\sigma_x\sigma_m-f_y\sigma_m\sigma_y\sigma_m}{\omega_n^2-\mu^2-2i\omega_n \mu+f_x^2+f_y^2}\psi_{i,<}(\bm k_1,\omega_{n}).
\end{align}
\end{small}
In the following, we consider the low frequency limit of the external leg, i.e., $\omega_n\to 0$. Define 
\begin{equation}
\mathcal{G}_1^{0}=\int\frac{\Lambda^3{\rm sin}\theta d\theta d\phi}{(2\pi)^3}\frac{1}{f_x^2(\Lambda,\theta,\phi)+f_y^2(\Lambda,\theta,\phi)-\mu^2},\\
\end{equation}
\begin{equation}
\mathcal{G}_1^{x}=\int\frac{\Lambda^3{\rm sin}\theta d\theta d\phi}{(2\pi)^3}\frac{f_x(\Lambda,\theta,\phi)}{f_x^2(\Lambda,\theta,\phi)+f_y^2(\Lambda,\theta,\phi)-\mu^2},\\
\label{G1x}
\end{equation}
\begin{equation}
\mathcal{G}_1^{y}=\int\frac{\Lambda^3{\rm sin}\theta d\theta d\phi}{(2\pi)^3}\frac{f_y(\Lambda,\theta,\phi)}{f_x^2(\Lambda,\theta,\phi)+f_y^2(\Lambda,\theta,\phi)-\mu^2},\\
\label{G1y}
\end{equation}
where $k_x=\Lambda {\rm cos}\phi{\rm sin}\theta$, $k_y=\Lambda {\rm sin}\phi{\rm sin}\theta$, and $k_z=\Lambda {\rm cos}\theta$. Then we have
\begin{align}
\nonumber
\delta S_{0,<}=&\frac{1}{\beta}\sum_{n}\int_< \frac{d^3{\bm k_1}}{(2\pi)^3} \sum\limits_{i=1}^{N} {\bar\psi_{i,<}}(\bm k_1,\omega_n)[(-i\omega_n-\mu)(\Delta_0+\Delta_x+\Delta_y+\Delta_z)\mathcal G_1^0dl\\
&+\sigma_x(-\Delta_0-\Delta_x+\Delta_y+\Delta_z)\mathcal G_1^xdl+\sigma_y(-\Delta_0+\Delta_x-\Delta_y+\Delta_z)\mathcal G_1^ydl]\psi_{i,<}(\bm k_1,\omega_{n}).
\end{align}
\begin{figure}[t]
	\centering
	\includegraphics[width=0.9\textwidth]{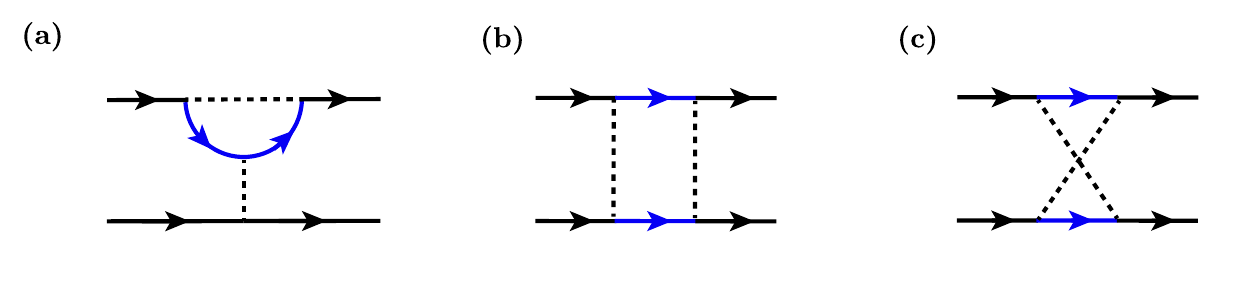}
	\caption{Diagrams for the second order perturbation.}
	\label{fig_scdorder}
\end{figure}
To the second order in $\Delta_m$, three types of diagrams should be considered as shown in Fig. \ref{fig_scdorder}. $S_{1,>}$ modifies $S_{1,<}$ as
\begin{align}
			\delta S_{1,<}=-\frac12 (\langle S_{1,>}^2\rangle_{0,>}-\langle S_{1,>}\rangle_{0,>}^2).
		\end{align}  
		The contribution from Fig. \ref{fig_scdorder} (a) is
		\begin{small}
		\begin{align}
			\nonumber
			&[\delta S_{1,<}]_{(a)}=-\frac12 \times2\times2\times2 \times\frac{1}{\beta^2}\sum_{n,n^\prime}\int\frac{d^3\bm k_1}{(2\pi)^3}\frac{d^3\bm k_2}{(2\pi)^3}\frac{d^3\bm k_3}{(2\pi)^3}\frac{d^3\bm k_4}{(2\pi)^3}\frac{d^3\bm k}{(2\pi)^3}(2\pi)^3\delta({\bm  k_1+\bm k_3-\bm k_2-\bm k_4})\sum_{i,j=1}^{N}\sum_{m,m^\prime=0,x,y,z}\frac{\Delta_m\Delta_{m^\prime}}{4}\\ \nonumber
			&\langle\bar\psi_{i,\textless}(\bm{k}_{1},\omega_n)\sigma_m\psi_{i,\textless}(\bm{k}_{2},\omega_n)\bar\psi_{j,>}(\bm{k}-\bm{q},\omega_{n^\prime})\sigma_m\psi_{j,>}(\bm{k},\omega_{n^\prime})\bar\psi_{j,>}(\bm{k},\omega_{n^\prime})\sigma_{m^\prime}\psi_{j,<}(\bm{k}_{4},\omega_{n^\prime})\bar\psi_{j,<}(\bm{k}_3,\omega_{n^\prime})\sigma_{m^\prime}\psi_{j,>}(\bm{k}-\bm{q},\omega_{n^\prime})\rangle_{0,>} \\ \nonumber
			&=-\frac12 \times2\times2\times2 \times\frac{1}{\beta^2}\sum_{n,n^\prime}\int\frac{d^3\bm k_1}{(2\pi)^3}\frac{d^3\bm k_2}{(2\pi)^3}\frac{d^3\bm k_3}{(2\pi)^3}\frac{d^3\bm k_4}{(2\pi)^3}\frac{d^3\bm k}{(2\pi)^3}(2\pi)^3\delta({\bm  k_1+\bm k_3-\bm k_2-\bm k_4})\sum_{i,j=1}^{N}\sum_{m,m^\prime=0,x,y,z}\frac{\Delta_m\Delta_{m^\prime}}{4}\\
			&\bar\psi_{i,\textless}(\bm{k}_{1},\omega_n)\sigma_m\psi_{i,\textless}(\bm{k}_{2},\omega_n)\bar\psi_{j,<}(\bm{k}_3,\omega_{n^\prime})\sigma_{m^\prime}[G_0(\bm k -\bm q,\omega_{n^\prime})]\sigma_m[G_0(\bm k,\omega_{n^\prime})]\sigma_{m^\prime}\psi_{j,<}(\bm{k}_{4},\omega_{n^\prime}),
		\end{align}
	\end{small}
		\\
		where $\bm q=\bm k_1-\bm k_2$ is the transfered momentum. Note that the multiplications of $2$'s represent the symmetry factor of the Feynman diagrams. The contribution from Fig. \ref{fig_scdorder} (b) is
		\begin{small}
		\begin{align}
			\nonumber
			&[\delta S_{1,<}]_{(b)}=-\frac12 \times2\times2 \times\frac{1}{\beta^2}\sum_{n,n^\prime}\int\frac{d^3\bm k_1}{(2\pi)^3}\frac{d^3\bm k_2}{(2\pi)^3}\frac{d^3\bm k_3}{(2\pi)^3}\frac{d^3\bm k_4}{(2\pi)^3}\frac{d^3\bm k}{(2\pi)^3}(2\pi)^3\delta({\bm  k_1+\bm k_3-\bm k_2-\bm k_4})\sum_{i,j=1}^{N}\sum_{m,m^\prime=0,x,y,z}\frac{\Delta_m\Delta_{m^\prime}}{4}\\ \nonumber
			&\langle\bar\psi_{i,\textless}(\bm{k}_{1},\omega_n)\sigma_m\psi_{i,>}(\bm{k},\omega_n)\bar\psi_{j,<}(\bm{k}_3,\omega_{n^\prime})\sigma_m\psi_{j,>}(\bm{q}-\bm{k},\omega_{n^\prime})\bar\psi_{i,>}(\bm{k},\omega_n)\sigma_{m^\prime}\psi_{i,<}(\bm{k}_{2},\omega_n)\bar\psi_{j,>}(\bm{q}-\bm{k},\omega_{n^\prime})\sigma_{m^\prime}\psi_{j,<}(\bm{k}_4,\omega_{n^\prime})\rangle_{0,>} \\ \nonumber
			&=-\frac12 \times2\times2\times\frac{1}{\beta^2}\sum_{n,n^\prime}\int\frac{d^3\bm k_1}{(2\pi)^3}\frac{d^3\bm k_2}{(2\pi)^3}\frac{d^3\bm k_3}{(2\pi)^3}\frac{d^3\bm k_4}{(2\pi)^3}\frac{d^3\bm k}{(2\pi)^3}(2\pi)^3\delta({\bm  k_1+\bm k_3-\bm k_2-\bm k_4})\sum_{i,j=1}^{N}\sum_{m,m^\prime=0,x,y,z}\frac{\Delta_m\Delta_{m^\prime}}{4}\\
			&\bar\psi_{i,\textless}(\bm{k}_{1},\omega_n)\sigma_m[G_0(\bm k,\omega_{n})]\sigma_{m^\prime}\psi_{i,<}(\bm{k}_{2},\omega_n)\bar\psi_{j,<}(\bm{k}_3,\omega_{n^\prime})\sigma_m[G_0(\bm{q}-\bm{k},\omega_{n^\prime})]\sigma_{m^\prime}\psi_{j,<}(\bm{k}_4,\omega_{n^\prime}),
		\end{align}
	\end{small}
		where $\bm q=\bm k_1+\bm k_3=\bm k_2+\bm k_4$.
		
		The contribution from Fig. \ref{fig_scdorder} (c) is
		\begin{small}
		\begin{align}
			\nonumber
			&[\delta S_{1,<}]_{(c)}=-\frac12 \times2\times2 \times\frac{1}{\beta^2}\sum_{n,n^\prime}\int\frac{d^3\bm k_1}{(2\pi)^3}\frac{d^3\bm k_2}{(2\pi)^3}\frac{d^3\bm k_3}{(2\pi)^3}\frac{d^3\bm k_4}{(2\pi)^3}\frac{d^3\bm k}{(2\pi)^3}(2\pi)^3\delta({\bm  k_1+\bm k_3-\bm k_2-\bm k_4})\sum_{i,j=1}^{N}\sum_{m,m^\prime=0,x,y,z}\frac{\Delta_m\Delta_{m^\prime}}{4}\\ \nonumber
			&\langle\bar\psi_{i,\textless}(\bm{k}_{1},\omega_n)\sigma_m\psi_{i,>}(\bm{k},\omega_n)\bar\psi_{j,>}(\bm{k}-
			\bm q,\omega_{n^\prime})\sigma_m\psi_{j,<}(\bm{k}_4,\omega_{n^\prime})\bar\psi_{i,>}(\bm{k},\omega_n)\sigma_{m^\prime}\psi_{i,<}(\bm{k}_{2},\omega_n)\bar\psi_{j,<}(\bm k_3,\omega_{n^\prime})\sigma_{m^\prime}\psi_{j,>}(\bm{k}-\bm q,\omega_{n^\prime})\rangle_{0,>} \\ \nonumber
			&=-\frac12 \times2\times2\times\frac{1}{\beta^2}\sum_{n,n^\prime}\int\frac{d^3\bm k_1}{(2\pi)^3}\frac{d^3\bm k_2}{(2\pi)^3}\frac{d^3\bm k_3}{(2\pi)^3}\frac{d^3\bm k_4}{(2\pi)^3}\frac{d^3\bm k}{(2\pi)^3}(2\pi)^3\delta({\bm  k_1+\bm k_3-\bm k_2-\bm k_4})\sum_{i,j=1}^{N}\sum_{m,m^\prime=0,x,y,z}\frac{\Delta_m\Delta_{m^\prime}}{4}\\
			&\bar\psi_{i,\textless}(\bm{k}_{1},\omega_n)\sigma_m[G_0(\bm k,\omega_{n})]\sigma_{m^\prime}\psi_{i,<}(\bm{k}_{2},\omega_n)\bar\psi_{j,<}(\bm{k}_3,\omega_{n^\prime})\sigma_{m^\prime}[G_0(\bm{k}-\bm{q},\omega_{n^\prime})]\sigma_m\psi_{j,<}(\bm{k}_4,\omega_{n^\prime}),
		\end{align}
	\end{small}
		where $\bm q=\bm k_1-\bm k_4$.
		
		We take the $\bm q$'s in the above relations to be 0 for that we only consider external legs with small momentum, and $\omega_n=0$ in the static limit. Moreover, the terms that emerge in $\delta S_{1,<}$ but do not appear in the original $S_{1,<}$ are omitted because they are of the order $O(\Delta^2)$ and break the constraint $\langle V_m({\bm r})V_n({\bm r^\prime}) \rangle=\delta_{mn}\delta({\bm r}-{\bm r^\prime})$. Therefore, the modification of $S_{1,<}$ from the above diagrams can be summarized as
		
		\begin{align}
			\nonumber
			\delta S_{1,<}=&-\frac{1}{\beta^2}\sum_{n,n^\prime}\int\frac{d^3\bm k_1}{(2\pi)^3}\frac{d^3\bm k_2}{(2\pi)^3}\frac{d^3\bm k_3}{(2\pi)^3}\frac{d^3\bm k_4}{(2\pi)^3}(2\pi)^3\delta({\bm  k_1+\bm k_3-\bm k_2-\bm k_4})\sum_{i,j=1}^{N}\frac12\times\\\nonumber
			&\{\bar\psi_{i,\textless}(\bm{k}_{1},\omega_n)\sigma_0\psi_{i,<}(\bm{k}_{2},\omega_n)\bar\psi_{j,<}(\bm{k}_3,\omega_{n^\prime})\sigma_0\psi_{j,<}(\bm{k}_4,\omega_{n^\prime})\\ \nonumber
			&[2\Delta_0(\Delta_0+\Delta_x+\Delta_y+\Delta_z)(\mu^2\mathcal G^{00}_2+\mathcal G^{xx}_2+\mathcal G^{yy}_2)+2(\Delta_0\Delta_x-\Delta_y\Delta_z)\mathcal G^{x\bar x}_2+2(\Delta_0\Delta_y-\Delta_x\Delta_z)\mathcal G^{y\bar y}_2\\\nonumber
			&+\mu^2(\mathcal G^{0 0}_2+\mathcal G^{0\bar 0}_2)(\Delta_0^2+\Delta_x^2+\Delta_y^2+\Delta_z^2)+2(\Delta_0\Delta_x+\Delta_y\Delta_z)\mathcal G^{xx}_2+2(\Delta_0\Delta_y+\Delta_x\Delta_z)\mathcal G^{yy}_2]dl\\\nonumber
			&+\bar\psi_{i,\textless}(\bm{k}_{1},\omega_n)\sigma_x\psi_{i,<}(\bm{k}_{2},\omega_n)\bar\psi_{j,<}(\bm{k}_3,\omega_{n^\prime})\sigma_x\psi_{j,<}(\bm{k}_4,\omega_{n^\prime})\\ \nonumber
			&[2\Delta_x(\Delta_0+\Delta_x-\Delta_y-\Delta_z)(\mu^2\mathcal G^{00}_2+\mathcal G^{xx}_2-\mathcal G^{yy}_2)+(\Delta_0^2+\Delta_x^2+\Delta_y^2+\Delta_z^2)\mathcal G^{x\bar x}_2+2(\Delta_x\Delta_y-\Delta_0\Delta_z)\mathcal G^{y\bar y}_2\\\nonumber
			&+2(\Delta_0\Delta_x-\Delta_y\Delta_z)\mu^2\mathcal G^{0\bar 0}_2+2(\Delta_0\Delta_x+\Delta_y\Delta_z)\mu^2\mathcal G^{00}_2+(\Delta_0^2+\Delta_x^2+\Delta_y^2+\Delta_z^2)\mathcal G^{xx}_2+2(\Delta_0\Delta_z+\Delta_x\Delta_y)\mathcal G^{yy}_2\\\nonumber
			&+4\Delta_x(\Delta_0+\Delta_x+\Delta_y+\Delta_z)\mu\mathcal G^{0x}_2]dl\\\nonumber
			&+\bar\psi_{i,\textless}(\bm{k}_{1},\omega_n)\sigma_y\psi_{i,<}(\bm{k}_{2},\omega_n)\bar\psi_{j,<}(\bm{k}_3,\omega_{n^\prime})\sigma_y\psi_{j,<}(\bm{k}_4,\omega_{n^\prime})\\ \nonumber
			&[2\Delta_y(\Delta_0+\Delta_y-\Delta_x-\Delta_z)(\mu^2\mathcal G^{00}_2+\mathcal G^{yy}_2-\mathcal G^{xx}_2)+(\Delta_0^2+\Delta_x^2+\Delta_y^2+\Delta_z^2)\mathcal G^{y\bar y}_2+2(\Delta_x\Delta_y-\Delta_0\Delta_z)\mathcal G^{x\bar x}_2\\\nonumber
			&+2(\Delta_0\Delta_y-\Delta_x\Delta_z)\mu^2\mathcal G^{0\bar 0}_2+2(\Delta_0\Delta_y+\Delta_x\Delta_z)\mu^2\mathcal G^{00}_2+(\Delta_0^2+\Delta_x^2+\Delta_y^2+\Delta_z^2)\mathcal G^{yy}_2+2(\Delta_0\Delta_z+\Delta_x\Delta_y)\mathcal G^{xx}_2\\\nonumber
			&+4\Delta_y(\Delta_0+\Delta_x+\Delta_y+\Delta_z)\mu\mathcal G^{0y}_2]dl\\\nonumber
			&+\bar\psi_{i,\textless}(\bm{k}_{1},\omega_n)\sigma_z\psi_{i,<}(\bm{k}_{2},\omega_n)\bar\psi_{j,<}(\bm{k}_3,\omega_{n^\prime})\sigma_z\psi_{j,<}(\bm{k}_4,\omega_{n^\prime})\\ \nonumber
			&[2\Delta_z(-\Delta_0+\Delta_x+\Delta_y-\Delta_z)(-\mu^2\mathcal G^{00}_2+\mathcal G^{xx}_2+\mathcal G^{yy}_2)+2(\Delta_x\Delta_z-\Delta_0\Delta_y)\mathcal G^{x\bar x}_2+2(\Delta_y\Delta_z-\Delta_0\Delta_x)\mathcal G^{y\bar y}_2\\
			&+2(\Delta_0\Delta_z-\Delta_x\Delta_y)\mu^2\mathcal G^{0\bar 0}_2+2(\Delta_0\Delta_z+\Delta_x\Delta_y)\mu^2\mathcal G^{0 0}_2+2(\Delta_0\Delta_y+\Delta_x\Delta_z)\mathcal G^{xx}_2+2(\Delta_0\Delta_x+\Delta_y\Delta_z)\mathcal G^{yy}_2]dl\}.
		\end{align}
		
		Here
		\begin{equation}
			\mathcal{G}_2^{00}=\int\frac{\Lambda^3{\rm sin}\theta d\theta d\phi}{(2\pi)^3}\frac{1}{[f_x^2(\Lambda,\theta,\phi)+f_y^2(\Lambda,\theta,\phi)-\mu^2]^2},
		\end{equation}
		\begin{equation}
			\mathcal{G}_2^{0x}=\int\frac{\Lambda^3{\rm sin}\theta d\theta d\phi}{(2\pi)^3}\frac{f_x(\Lambda,\theta,\phi)}{[f_x^2(\Lambda,\theta,\phi)+f_y^2(\Lambda,\theta,\phi)-\mu^2]^2},
		\end{equation}
		\begin{equation}
			\mathcal{G}_2^{0y}=\int\frac{\Lambda^3{\rm sin}\theta d\theta d\phi}{(2\pi)^3}\frac{f_y(\Lambda,\theta,\phi)}{[f_x^2(\Lambda,\theta,\phi)+f_y^2(\Lambda,\theta,\phi)-\mu^2]^2},
		\end{equation}
		\begin{equation}
			\mathcal{G}_2^{xx}=\int\frac{\Lambda^3{\rm sin}\theta d\theta d\phi}{(2\pi)^3}\frac{f_x^2(\Lambda,\theta,\phi)}{[f_x^2(\Lambda,\theta,\phi)+f_y^2(\Lambda,\theta,\phi)-\mu^2]^2},
			\label{G2xx}
		\end{equation}
		\begin{equation}
			\mathcal{G}_2^{yy}=\int\frac{\Lambda^3{\rm sin}\theta d\theta d\phi}{(2\pi)^3}\frac{f_y^2(\Lambda,\theta,\phi)}{[f_x^2(\Lambda,\theta,\phi)+f_y^2(\Lambda,\theta,\phi)-\mu^2]^2},
				\label{G2yy}
		\end{equation}
		\begin{equation}
			\mathcal{G}_2^{0\bar 0}=\int\frac{\Lambda^3{\rm sin}\theta d\theta d\phi}{(2\pi)^3}\frac{1}{[f_x^2(\Lambda,\theta,\phi)+f_y^2(\Lambda,\theta,\phi)-\mu^2]\cdot[f_x^2(\Lambda,\pi-\theta,\pi+\phi)+f_y^2(\Lambda,\pi-\theta,\pi+\phi)-\mu^2]},
		\end{equation}
		\begin{equation}
			\mathcal{G}_2^{x\bar x}=\int\frac{\Lambda^3{\rm sin}\theta d\theta d\phi}{(2\pi)^3}\frac{f_x(\Lambda,\theta,\phi)f_x(\Lambda,\pi-\theta,\pi+\phi)}{[f_x^2(\Lambda,\theta,\phi)+f_y^2(\Lambda,\theta,\phi)-\mu^2]\cdot[f_x^2(\Lambda,\pi-\theta,\pi+\phi)+f_y^2(\Lambda,\pi-\theta,\pi+\phi)-\mu^2]},
		\end{equation}
		\begin{equation}
			\mathcal{G}_2^{y\bar y}=\int\frac{\Lambda^3{\rm sin}\theta d\theta d\phi}{(2\pi)^3}\frac{f_y(\Lambda,\theta,\phi)f_y(\Lambda,\pi-\theta,\pi+\phi)}{[f_x^2(\Lambda,\theta,\phi)+f_y^2(\Lambda,\theta,\phi)-\mu^2]\cdot[f_x^2(\Lambda,\pi-\theta,\pi+\phi)+f_y^2(\Lambda,\pi-\theta,\pi+\phi)-\mu^2]}.
		\end{equation}
		\section{RG step 2: rescale}
		After collecting all one-loop corrections, we perform the rescaling of space-time coordinates as 
		\begin{align}
			\nonumber
			\bm r &\to \bm r^\prime=e^{-dl} \bm r,\nonumber\\
\bm k &\to \bm k^\prime=e^{dl} \bm k,\nonumber\\
\tau &\to \tau^\prime = e^{-\kappa dl}\tau,\nonumber\\
\omega &\to \omega^\prime = e^{\kappa dl}\omega.
	    \end{align}
Then we introduce the renormalization constants
\begin{align} 
	\psi=Z_{\psi}^{-1/2}\psi^{\prime},\quad
	v_{p/q}=Z_{v_{p/q}}^{-1}v_{p/q}^{\prime},\quad\mu=Z_{\mu}^{-1}\mu^{\prime},\quad\lambda_i=Z_{\lambda_i}^{-1}\lambda_i^{\prime},\quad m_i=Z_{m_i}^{-1}m_i^{\prime},\quad \Delta_i=Z_{\Delta_i}^{-1}\Delta_i^{\prime}.
\end{align}	
		The action in terms of the new variables reads (from now on we relabel $X^\prime$'s to $X$'s for simplicity).
		\begin{small}
		\begin{align}
			\nonumber
			S=&\int \frac{d^3{\bm k}}{(2\pi)^3}\frac{d\omega}{2\pi} e^{-(3+\kappa)dl}\sum\limits_{i=1}^{N}\frac{1}{Z_{\psi}}\bar{\psi}_i({\bm k},\omega)\{-i\omega e^{-\kappa dl}[1+(\Delta_0+\Delta_x+\Delta_y+\Delta_z)\mathcal G_1^0 dl]-\frac{\mu}{Z_{\mu}}[1+(\Delta_0+\Delta_x+\Delta_y+\Delta_z)\mathcal G_1^0 dl]\\ \nonumber
			&+[\frac{v_p}{Z_{v_p}}e^{-pdl}{\rm Re}(k_{x}+ik_{y})^{p}+\frac{v_q}{Z_{v_q}}{\rm Re}(e^{-dl}k_{z}+i(\frac{\lambda_1}{Z_{\lambda_1}}-e^{-2dl}\frac{\lambda_2}{Z_{\lambda_2}}k^2))^{q}+\frac{m_1}{Z_{m_1}}+(-\Delta_0-\Delta_x+\Delta_y+\Delta_z)\mathcal G_1^xdl]\sigma_x\\ \nonumber
			&+[\frac{v_p}{Z_{v_p}}e^{-pdl}{\rm Im}(k_{x}+ik_{y})^{p}+\frac{v_q}{Z_{v_q}}{\rm Im}(e^{-dl}k_{z}+i(\frac{\lambda_1}{Z_{\lambda_1}}-e^{-2dl}\frac{\lambda_2}{Z_{\lambda_2}}k^2))^{q}+\frac{m_2}{Z_{m_2}}+(-\Delta_0+\Delta_x-\Delta_y+\Delta_z)\mathcal G_1^ydl]\sigma_y\}\psi_i({\bm k},\omega)\\ \nonumber
			&-\frac{1}{2}\int \frac{d^3\bm k_1}{(2\pi)^3}\frac{d^3\bm k_2}{(2\pi)^3}\frac{d^3\bm k_3}{(2\pi)^3}\frac{d^3\bm k_4}{(2\pi)^3}(2\pi)^3\delta({\bm  k_1+\bm k_3-\bm k_2-\bm k_4})\frac{d\omega^\prime }{2\pi}\frac{d\omega}{2\pi} e^{-(2\kappa+9)dl}\times\\ \nonumber 
			&\sum\limits_{i,j=1}^{N}\{\frac{\Delta_0}{Z_{\psi}^2Z_{\Delta_0}}\bar{\psi_{i}}({\bm k}_1,\omega)\sigma_0\psi_{i}({\bm k}_2,\omega)\bar{\psi}_{j}({\bm k}_3,\omega^\prime)\sigma_0\psi_{j}({\bm k}_4,\omega^\prime)\\ \nonumber
			&[1+[2(\Delta_0+\Delta_x+\Delta_y+\Delta_z)(\mu^2\mathcal G^{00}_2+\mathcal G^{xx}_2+\mathcal G^{yy}_2)+2(\Delta_x\Delta_0-\Delta_y\Delta_z)/\Delta_0\mathcal G^{x\bar x}_2+2(\Delta_y\Delta_0-\Delta_x\Delta_z)/\Delta_0\mathcal G^{y\bar y}_2\\ \nonumber
			&+\mu^2(\mathcal G^{0 0}_2+\mathcal G^{0\bar 0}_2)(\Delta_0^2+\Delta_x^2+\Delta_y^2+\Delta_z^2)/\Delta_0+2(\Delta_x\Delta_0+\Delta_y\Delta_z)/\Delta_0\mathcal G^{xx}_2+2(\Delta_y\Delta_0+\Delta_x\Delta_z)/\Delta_0\mathcal G^{yy}_2]dl]\\ \nonumber
			&+\frac{\Delta_x}{Z_{\psi}^2Z_{\Delta_x}}\bar{\psi_{i}}({\bm k}_1,\omega)\sigma_x\psi_{i}({\bm k}_2,\omega)\bar{\psi}_{j}({\bm k}_3,\omega^\prime)\sigma_x\psi_{j}({\bm k}_4,\omega^\prime)\\ \nonumber
			&[1+[2(\Delta_0+\Delta_x-\Delta_y-\Delta_z)(\mu^2\mathcal G^{0 0}_2+\mathcal G^{xx}_2-\mathcal G^{yy}_2)+(\Delta_0^2+\Delta_x^2+\Delta_y^2+\Delta_z^2)/\Delta_x\mathcal G^{x\bar x}_2+2(\Delta_x\Delta_y-\Delta_0\Delta_z)/\Delta_x\mathcal G^{y\bar y}_2\\ \nonumber
			&+2(\Delta_0\Delta_x-\Delta_y\Delta_z)/\Delta_x\mu^2\mathcal G^{0\bar 0}_2+2(\Delta_0\Delta_x+\Delta_y\Delta_z)/\Delta_x\mu^2\mathcal G^{00}_2+4(\Delta_0+\Delta_x+\Delta_y+\Delta_z)\mu\mathcal G^{0x}_2\\\nonumber
			&+(\Delta_0^2+\Delta_x^2+\Delta_y^2+\Delta_z^2)/\Delta_x\mathcal G^{xx}_2+2(\Delta_0\Delta_z+\Delta_x\Delta_y)/\Delta_x\mathcal G^{yy}_2]dl]\\ \nonumber
			&+\frac{\Delta_y}{Z_{\psi}^2Z_{\Delta_y}}\bar{\psi_{i}}({\bm k}_1,\omega)\sigma_y\psi_{i}({\bm k}_2,\omega)\bar{\psi}_{j}({\bm k}_3,\omega^\prime)\sigma_y\psi_{j}({\bm k}_4,\omega^\prime)\\ \nonumber
			&[1+[2(\Delta_0+\Delta_y-\Delta_x-\Delta_z)(\mu^2\mathcal G^{0 0}_2+\mathcal G^{yy}_2-\mathcal G^{xx}_2)+(\Delta_0^2+\Delta_y^2+\Delta_x^2+\Delta_z^2)/\Delta_y\mathcal G^{y\bar y}_2+2(\Delta_x\Delta_y-\Delta_0\Delta_z)/\Delta_y\mathcal G^{x\bar x}_2\\ \nonumber
			&+2(\Delta_0\Delta_y-\Delta_x\Delta_z)/\Delta_y\mu^2\mathcal G^{0\bar 0}_2+2(\Delta_0\Delta_y+\Delta_x\Delta_z)/\Delta_y\mu^2\mathcal G^{00}_2+4(\Delta_0+\Delta_x+\Delta_y+\Delta_z)\mu\mathcal G^{0y}_2\\\nonumber
			&+(\Delta_0^2+\Delta_x^2+\Delta_y^2+\Delta_z^2)/\Delta_y\mathcal G^{yy}_2+2(\Delta_0\Delta_z+\Delta_x\Delta_y)/\Delta_y\mathcal G^{xx}_2]dl]\\\nonumber
			&+\frac{\Delta_z}{Z_{\psi}^2Z_{\Delta_z}}\bar{\psi_{i}}({\bm k}_1,\omega)\sigma_z\psi_{i}({\bm k}_2,\omega)\bar{\psi}_{j}({\bm k}_3,\omega^\prime)\sigma_z\psi_{j}({\bm k}_4,\omega^\prime)\\\nonumber 
			&[1+[2(-\Delta_0+\Delta_x+\Delta_y-\Delta_z)(-\mu^2\mathcal G^{00}_2+\mathcal G^{xx}_2+\mathcal G^{yy}_2)+2(\Delta_x\Delta_z-\Delta_0\Delta_y)/\Delta_z\mathcal G^{x\bar x}_2+2(\Delta_y\Delta_z-\Delta_0\Delta_x)/\Delta_z\mathcal G^{y\bar y}_2\\ \nonumber
			&+2(\Delta_0\Delta_z-\Delta_x\Delta_y)/\Delta_z\mu^2\mathcal G^{0\bar 0}_2+2(\Delta_0\Delta_z+\Delta_x\Delta_y)/\Delta_z\mu^2\mathcal G^{0 0}_2\\
			&+2(\Delta_0\Delta_y+\Delta_x\Delta_z)/\Delta_z\mathcal G^{xx}_2+2(\Delta_0\Delta_x+\Delta_y\Delta_z)/\Delta_z\mathcal G^{yy}_2]dl]\}.
		\end{align}
	\end{small}

		\section{RG step 3: renormalization and the RG equations}
		By requiring that $S$ is of the same form of the non-renormalized case but with rescaled space, time, and parameters, we obtain the renormalization constants as
		\begin{small}
		\begin{align}
			\nonumber
			Z_{\psi}=&1+[-3-2\kappa+(\Delta_0+\Delta_x+\Delta_y+\Delta_z)\mathcal G_1^0 ]dl,\quad Z_{\mu}=1+\kappa dl,\\\nonumber  Z_{v_p}=&1+[\kappa-p-(\Delta_0+\Delta_x+\Delta_y+\Delta_z)\mathcal G_1^0 ]dl,\quad Z_{v_q}=1+[\kappa-q-(\Delta_0+\Delta_x+\Delta_y+\Delta_z)\mathcal G_1^0 ]dl,\\ \nonumber
			Z_{\lambda_1}=&1+dl,\quad Z_{\lambda_2}=1-dl,\\ \nonumber
			Z_{m_1}=&1+[\kappa-(\Delta_0+\Delta_x+\Delta_y+\Delta_z)\mathcal G_1^0+(-\Delta_0-\Delta_x+\Delta_y+\Delta_z)\mathcal G_1^x/m_1]dl,\\ \nonumber
			Z_{m_2}=&1+[\kappa-(\Delta_0+\Delta_x+\Delta_y+\Delta_z)\mathcal G_1^0+(-\Delta_0+\Delta_x-\Delta_y+\Delta_z)\mathcal G_1^y/m_2]dl,\\ \nonumber
			Z_{\Delta_0}=&1+[2\kappa-3-2(\Delta_0+\Delta_x+\Delta_y+\Delta_z)\mathcal G_1^0+2(\Delta_0+\Delta_x+\Delta_y+\Delta_z)(\mu^2\mathcal G^{00}_2+\mathcal G^{xx}_2+\mathcal G^{yy}_2)\\\nonumber&+2(\Delta_x\Delta_0-\Delta_y\Delta_z)/\Delta_0\mathcal G^{x\bar x}_2+2(\Delta_y\Delta_0-\Delta_x\Delta_z)/\Delta_0\mathcal G^{y\bar y}_2+\mu^2(\mathcal G^{0 0}_2+\mathcal G^{0\bar 0}_2)(\Delta_0^2+\Delta_x^2+\Delta_y^2+\Delta_z^2)/\Delta_0\\ \nonumber
			&+2(\Delta_x\Delta_0+\Delta_y\Delta_z)/\Delta_0\mathcal G^{xx}_2+2(\Delta_y\Delta_0+\Delta_x\Delta_z)/\Delta_0\mathcal G^{yy}_2]dl,\\ \nonumber
			Z_{\Delta_x}=&1+[2\kappa-3-2(\Delta_0+\Delta_x+\Delta_y+\Delta_z)\mathcal G_1^0+2(\Delta_0+\Delta_x-\Delta_y-\Delta_z)(\mu^2\mathcal G^{0 0}_2+\mathcal G^{xx}_2-\mathcal G^{yy}_2)+(\Delta_0^2+\Delta_x^2+\Delta_y^2+\Delta_z^2)/\Delta_x\mathcal G^{x\bar x}_2\\ \nonumber
			&+2(\Delta_x\Delta_y-\Delta_0\Delta_z)/\Delta_x\mathcal G^{y\bar y}_2+2(\Delta_0\Delta_x-\Delta_y\Delta_z)/\Delta_x\mu^2\mathcal G^{0\bar 0}_2+2(\Delta_0\Delta_x+\Delta_y\Delta_z)/\Delta_x\mu^2\mathcal G^{00}_2+4(\Delta_0+\Delta_x+\Delta_y+\Delta_z)\mu\mathcal G^{0x}_2\\\nonumber
			&+(\Delta_0^2+\Delta_x^2+\Delta_y^2+\Delta_z^2)/\Delta_x\mathcal G^{xx}_2+2(\Delta_0\Delta_z+\Delta_x\Delta_y)/\Delta_x\mathcal G^{yy}_2]dl,\\ \nonumber
			Z_{\Delta_y}=&1+[2\kappa-3-2(\Delta_0+\Delta_x+\Delta_y+\Delta_z)\mathcal G_1^0+2(\Delta_0+\Delta_y-\Delta_x-\Delta_z)(\mu^2\mathcal G^{0 0}_2+\mathcal G^{yy}_2-\mathcal G^{xx}_2)+(\Delta_0^2+\Delta_y^2+\Delta_x^2+\Delta_z^2)/\Delta_y\mathcal G^{y\bar y}_2\\ \nonumber
			&+2(\Delta_x\Delta_y-\Delta_0\Delta_z)/\Delta_y\mathcal G^{x\bar x}_2+2(\Delta_0\Delta_y-\Delta_x\Delta_z)/\Delta_y\mu^2\mathcal G^{0\bar 0}_2+2(\Delta_0\Delta_y+\Delta_x\Delta_z)/\Delta_y\mu^2\mathcal G^{00}_2+4(\Delta_0+\Delta_x+\Delta_y+\Delta_z)\mu\mathcal G^{0y}_2\\\nonumber
			&+(\Delta_0^2+\Delta_x^2+\Delta_y^2+\Delta_z^2)/\Delta_y\mathcal G^{yy}_2+2(\Delta_0\Delta_z+\Delta_x\Delta_y)/\Delta_y\mathcal G^{xx}_2]dl,\\\nonumber
			Z_{\Delta_z}=&1+[2\kappa-3-2(\Delta_0+\Delta_x+\Delta_y+\Delta_z)\mathcal G_1^0+2(\Delta_0-\Delta_x-\Delta_y+\Delta_z)(\mu^2\mathcal G^{00}_2-\mathcal G^{xx}_2-\mathcal G^{yy}_2)+2(\Delta_x\Delta_z-\Delta_0\Delta_y)/\Delta_z\mathcal G^{x\bar x}_2\\ \nonumber
			&+2(\Delta_y\Delta_z-\Delta_0\Delta_x)/\Delta_z\mathcal G^{y\bar y}_2+2(\Delta_0\Delta_z-\Delta_x\Delta_y)/\Delta_z\mu^2\mathcal G^{0\bar 0}_2+2(\Delta_0\Delta_z+\Delta_x\Delta_y)/\Delta_z\mu^2\mathcal G^{0 0}_2\\
			&+2(\Delta_0\Delta_y+\Delta_x\Delta_z)/\Delta_z\mathcal G^{xx}_2+2(\Delta_0\Delta_x+\Delta_y\Delta_z)/\Delta_z\mathcal G^{yy}_2]dl.
		\end{align}
	\end{small}
		The renormalization group equations for the parameters are obtained according to 
		\begin{equation}
			\frac{d{\rm ln}X}{dl}=\frac{dZ_X}{dl}\Big|_{l\to 0}.
		\end{equation}  
		Here, $X$ is some parameter but not the field $\psi$. Then we have the RG equations
		\begin{small}
		\begin{align}
			\nonumber
			\frac{d\mu}{dl}=&\kappa\mu,\quad\frac{dv_p}{dl}=v_p[\kappa-p-(\Delta_0+\Delta_x+\Delta_y+\Delta_z)\mathcal G_1^0 ],\quad\frac{dv_q}{dl}=v_q[\kappa-q-(\Delta_0+\Delta_x+\Delta_y+\Delta_z)\mathcal G_1^0 ],\\\nonumber 
			\frac{d\lambda_1}{dl}&=\lambda_1,\quad\frac{d\lambda_2}{dl}=-\lambda_2,\\ \nonumber
			\frac{dm_1}{dl}=&m_1[\kappa-(\Delta_0+\Delta_x+\Delta_y+\Delta_z)\mathcal G_1^0]+(-\Delta_0-\Delta_x+\Delta_y+\Delta_z)\mathcal G_1^x,\\ \nonumber
			\frac{dm_2}{dl}=&m_2[\kappa-(\Delta_0+\Delta_x+\Delta_y+\Delta_z)\mathcal G_1^0]+(-\Delta_0+\Delta_x-\Delta_y+\Delta_z)\mathcal G_1^y,\\ \nonumber
			\frac{d\Delta_0}{dl}=&\Delta_0[2\kappa-3-2(\Delta_0+\Delta_x+\Delta_y+\Delta_z)\mathcal G_1^0]+2\Delta_0(\Delta_0+\Delta_x+\Delta_y+\Delta_z)(\mu^2\mathcal G^{00}_2+\mathcal G^{xx}_2+\mathcal G^{yy}_2)\\\nonumber&+2(\Delta_x\Delta_0-\Delta_y\Delta_z)\mathcal G^{x\bar x}_2+2(\Delta_y\Delta_0-\Delta_x\Delta_z)/\mathcal G^{y\bar y}_2+\mu^2(\mathcal G^{0 0}_2+\mathcal G^{0\bar 0}_2)(\Delta_0^2+\Delta_x^2+\Delta_y^2+\Delta_z^2)\\ \nonumber
			&+2(\Delta_x\Delta_0+\Delta_y\Delta_z)\mathcal G^{xx}_2+2(\Delta_y\Delta_0+\Delta_x\Delta_z)\mathcal G^{yy}_2,\\ \nonumber
			\frac{d\Delta_x}{dl}=&\Delta_x[2\kappa-3-2(\Delta_0+\Delta_x+\Delta_y+\Delta_z)\mathcal G_1^0+2(\Delta_0+\Delta_x-\Delta_y-\Delta_z)(\mu^2\mathcal G^{0 0}_2+\mathcal G^{xx}_2-\mathcal G^{yy}_2)+4(\Delta_0+\Delta_x+\Delta_y+\Delta_z)\mu\mathcal G^{0x}_2]\\ \nonumber
			&+(\Delta_0^2+\Delta_x^2+\Delta_y^2+\Delta_z^2)\mathcal G^{x\bar x}_2+2(\Delta_x\Delta_y-\Delta_0\Delta_z)\mathcal G^{y\bar y}_2+2(\Delta_0\Delta_x-\Delta_y\Delta_z)\mu^2\mathcal G^{0\bar 0}_2+2(\Delta_0\Delta_x+\Delta_y\Delta_z)\mu^2\mathcal G^{00}_2\\\nonumber
			&+(\Delta_0^2+\Delta_x^2+\Delta_y^2+\Delta_z^2)\mathcal G^{xx}_2+2(\Delta_0\Delta_z+\Delta_x\Delta_y)\mathcal G^{yy}_2,\\ \nonumber
			\frac{d\Delta_y}{dl}=&\Delta_y[2\kappa-3-2(\Delta_0+\Delta_x+\Delta_y+\Delta_z)\mathcal G_1^0+2(\Delta_0+\Delta_y-\Delta_x-\Delta_z)(\mu^2\mathcal G^{0 0}_2+\mathcal G^{yy}_2-\mathcal G^{xx}_2)+4(\Delta_0+\Delta_x+\Delta_y+\Delta_z)\mu\mathcal G^{0y}_2]\\ \nonumber
			&+(\Delta_0^2+\Delta_y^2+\Delta_x^2+\Delta_z^2)\mathcal G^{y\bar y}_2+2(\Delta_x\Delta_y-\Delta_0\Delta_z)\mathcal G^{x\bar x}_2+2(\Delta_0\Delta_y-\Delta_x\Delta_z)\mu^2\mathcal G^{0\bar 0}_2+2(\Delta_0\Delta_y+\Delta_x\Delta_z)\mu^2\mathcal G^{00}_2\\\nonumber
			&+(\Delta_0^2+\Delta_x^2+\Delta_y^2+\Delta_z^2)\mathcal G^{yy}_2+2(\Delta_0\Delta_z+\Delta_x\Delta_y)\mathcal G^{xx}_2,\\\nonumber
			\frac{d\Delta_z}{dl}=&\Delta_z[2\kappa-3-2(\Delta_0+\Delta_x+\Delta_y+\Delta_z)\mathcal G_1^0+2(\Delta_0-\Delta_x-\Delta_y+\Delta_z)(\mu^2\mathcal G^{00}_2-\mathcal G^{xx}_2-\mathcal G^{yy}_2)]\\ \nonumber
			&+2(\Delta_x\Delta_z-\Delta_0\Delta_y)\mathcal G^{x\bar x}_2+2(\Delta_y\Delta_z-\Delta_0\Delta_x)\mathcal G^{y\bar y}_2+2(\Delta_0\Delta_z-\Delta_x\Delta_y)\mu^2\mathcal G^{0\bar 0}_2+2(\Delta_0\Delta_z+\Delta_x\Delta_y)\mu^2\mathcal G^{0 0}_2\\
			&+2(\Delta_0\Delta_y+\Delta_x\Delta_z)\mathcal G^{xx}_2+2(\Delta_0\Delta_x+\Delta_y\Delta_z)\mathcal G^{yy}_2.
		\end{align}
	\end{small}
		By requiring that the $k$-linear terms are scale-invariant, we obtain the dynamical exponent
		\begin{equation}
			\kappa=1+(\Delta_0+\Delta_x+\Delta_y+\Delta_z)\mathcal G_1^0.
		\end{equation} 
		The RG equations then become
\begin{align}
	           \nonumber
				\frac{d\mu}{dl}=&\mu[1+(\Delta_0+\Delta_x+\Delta_y+\Delta_z)\mathcal G_1^0 ],\quad\frac{dv_p}{dl}=(1-p)v_p,\quad
				\frac{dv_q}{dl}=(1-q)v_q,\quad\frac{d\lambda_1}{dl}=\lambda_1,\quad
				\frac{d\lambda_2}{dl}=-\lambda_2,\nonumber\\ 
				\frac{dm_1}{dl}=&m_1+(-\Delta_0-\Delta_x+\Delta_y+\Delta_z)\mathcal G_1^x,\quad \frac{dm_2}{dl}=m_2+(-\Delta_0+\Delta_x-\Delta_y+\Delta_z)\mathcal G_1^y,\nonumber\\ 
				\frac{d\Delta_0}{dl}=&-\Delta_0+2\Delta_0(\Delta_0+\Delta_x+\Delta_y+\Delta_z)(\mu^2\mathcal G^{00}_2+\mathcal G^{xx}_2+\mathcal G^{yy}_2)\nonumber\\&+2(\Delta_x\Delta_0-\Delta_y\Delta_z)\mathcal G^{x\bar x}_2+2(\Delta_y\Delta_0-\Delta_x\Delta_z)\mathcal G^{y\bar y}_2+\mu^2(\mathcal G^{0 0}_2+\mathcal G^{0\bar 0}_2)(\Delta_0^2+\Delta_x^2+\Delta_y^2+\Delta_z^2)\nonumber\\ 
				&+2(\Delta_x\Delta_0+\Delta_y\Delta_z)\mathcal G^{xx}_2+2(\Delta_y\Delta_0+\Delta_x\Delta_z)\mathcal G^{yy}_2,\nonumber\\ 
				\frac{d\Delta_x}{dl}=&-\Delta_x+2\Delta_x(\Delta_0+\Delta_x-\Delta_y-\Delta_z)(\mu^2\mathcal G^{0 0}_2+\mathcal G^{xx}_2-\mathcal G^{yy}_2)+4\Delta_x(\Delta_0+\Delta_x+\Delta_y+\Delta_z)\mu\mathcal G^{0x}_2\nonumber\\ 
				&+(\Delta_0^2+\Delta_x^2+\Delta_y^2+\Delta_z^2)\mathcal G^{x\bar x}_2+2(\Delta_x\Delta_y-\Delta_0\Delta_z)\mathcal G^{y\bar y}_2+2(\Delta_0\Delta_x-\Delta_y\Delta_z)\mu^2\mathcal G^{0\bar 0}_2+2(\Delta_0\Delta_x+\Delta_y\Delta_z)\mu^2\mathcal G^{00}_2\nonumber\\
				&+(\Delta_0^2+\Delta_x^2+\Delta_y^2+\Delta_z^2)\mathcal G^{xx}_2+2(\Delta_0\Delta_z+\Delta_x\Delta_y)\mathcal G^{yy}_2,\nonumber\\ 
				\frac{d\Delta_y}{dl}=&-\Delta_y+2\Delta_y(\Delta_0-\Delta_x+\Delta_y-\Delta_z)(\mu^2\mathcal G^{0 0}_2+\mathcal G^{yy}_2-\mathcal G^{xx}_2)+4\Delta_y(\Delta_0+\Delta_x+\Delta_y+\Delta_z)\mu\mathcal G^{0y}_2\nonumber\\ 
				&+(\Delta_0^2+\Delta_y^2+\Delta_x^2+\Delta_z^2)\mathcal G^{y\bar y}_2+2(\Delta_x\Delta_y-\Delta_0\Delta_z)\mathcal G^{x\bar x}_2+2(\Delta_0\Delta_y-\Delta_x\Delta_z)\mu^2\mathcal G^{0\bar 0}_2+2(\Delta_0\Delta_y+\Delta_x\Delta_z)\mu^2\mathcal G^{00}_2\nonumber\\
				&+(\Delta_0^2+\Delta_x^2+\Delta_y^2+\Delta_z^2)\mathcal G^{yy}_2+2(\Delta_0\Delta_z+\Delta_x\Delta_y)\mathcal G^{xx}_2,\nonumber\\ \nonumber
				\frac{d\Delta_z}{dl}=&-\Delta_z+2\Delta_z(\Delta_0-\Delta_x-\Delta_y+\Delta_z)(\mu^2\mathcal G^{00}_2-\mathcal G^{xx}_2-\mathcal G^{yy}_2)+2(\Delta_x\Delta_z-\Delta_0\Delta_y)\mathcal G^{x\bar x}_2+2(\Delta_y\Delta_z-\Delta_0\Delta_x)\mathcal G^{y\bar y}_2\\
				&+2(\Delta_0\Delta_z-\Delta_x\Delta_y)\mu^2\mathcal G^{0\bar 0}_2+2(\Delta_0\Delta_z+\Delta_x\Delta_y)\mu^2\mathcal G^{0 0}_2+2(\Delta_0\Delta_y+\Delta_x\Delta_z)\mathcal G^{xx}_2+2(\Delta_0\Delta_x+\Delta_y\Delta_z)\mathcal G^{yy}_2. \label{EqRGeqsT}
		\end{align}
In the absence of disorder, the parameters depends on $l$ in a simple manner as
		\begin{align}
			\nonumber
			\mu(l)=&\mu(l=0)e^{l},\quad v_p(l)=v_p(l=0)e^{(1-p)l},\quad v_q(l)=v_q(l=0)e^{(1-q)l},\\
			\lambda_1(l)=&\lambda_1(l=0)e^{l},\quad \lambda_2(l)=\lambda_2(l=0)e^{-l},\quad m_i(l)=m_i(l=0)e^{l}.
		\end{align} 			
		
		Equations.\,(6-8) in the main text can be obtained by setting $\mu=0$ for Eq.\,\ref{EqRGeqsT}. Specially, Eq.\,(7) or (8) in the main text are obtained by setting $\Delta_x=\Delta_y=\Delta_z=0$ or $\Delta_0=\Delta_x=\Delta_y=0$. In numerating Eq.\,(\ref{G1x}), (\ref{G1y}), (\ref{G2xx}), and (\ref{G2yy}), we set  
\begin{align}
	\nonumber
	f_x(\bm k)&=v_2[k_x^2-k_y^2+k_z^2-(\lambda_1-\lambda_2 k^2)^2]+m_1,\\
	f_y(\bm k)&=v_2[k_x k_y+k_z(\lambda_1-\lambda_2 k^2)]+m_2,
\end{align}
for the Hopf linked nodal-knots, and 
\begin{align}
	\nonumber
	f_x(\bm k)&=v_3[k_x^3-3k_x k_y^2+k_z^3-3k_z(\lambda_1-\lambda_2 k^2)^2]+m_1,\\
	f_y(\bm k)&=v_3[3k_x^2 k_y-k_y^3+3k_z^2(\lambda_1-\lambda_2 k^2)-(\lambda_1-\lambda_2 k^2)^3]+m_2,
\end{align}		
for the valknut nodal-knots.		
\section{Knot Wilson loop integrals}	
\begin{figure}[h]
	\centering
	\includegraphics[width=0.95\textwidth]{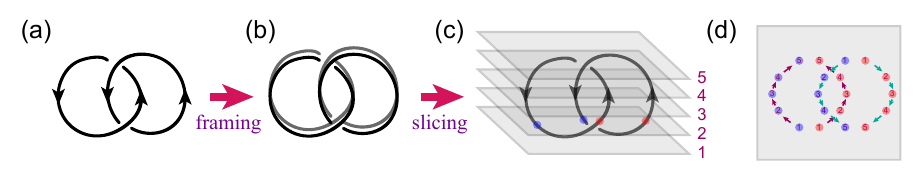}
	\caption{Schematics of the numerical method to calculate the knot Wilson loop integral. (a), the Hopf-linked nodal-knot. (b), to avoid the singularity of the Berry phase on the nodal lines, the integral paths are slightly shifted from the nodal lines (the procedure was also called framing). (c), in calculating the Wilson loop integral, the 1D line integral in 3D Brillouin zone is sliced into many 2D layers (schematically numered as 1-5). The directed integral paths intersects each layer at four points as shown by colored circles in (d). The colors (red or blue) of the circles denote the direction of the integral paths that pass the corresponding layer. The accumulated Berry phase can be obtained by counting the net phase changes of the intersection points (e.g. $\Delta \Phi=\Delta\Phi_{\rm red, 1\to 5}-\Delta\Phi_{\rm blue, 1\to 5}$).}
	\label{figRwilson}
\end{figure}	
We utilize the knot Wilson loop integral to characterize specific knot configurations to compensate for the limitations of solely using DOS markers to mark knot transition boundaries \cite{witten_quantum_1989,lian_chern-simons_2017}. The the knot Wilson loop gives rise to the total Berry phase accumulated by electrons as they traverse along the knotted nodal line and return to their original position. In the following, we first introduce the computational algorithm for the knot Wilson loop integral. Next, we present the results obtained through numerical calculations for systems with various configurations. Subsequently, we demonstrate the phase diagram determined jointly by DOS markers and the knot Berry phase. Finally, we discuss potential experimentally observables corresponding to the knot Wilson loop.

We take the Hopf-linked nodal-knot as the example to calculate the knot Wilson loop. The knot Wilson loop with $N$ oriented loops labeled by $L_1,\cdots,L_N$ (in the case of Hopf link, $N=2$ and the two linked loops are labeled by $L_1$ and $L_2$) can be obtained through 
\begin{eqnarray}
W(L_1,\cdots,L_N)=\frac{1}{\pi}\oint_{\bm{l}\in L_1,\cdots,L_N}\bm{A}(\bm{k})\cdot d\bm{l}.
\end{eqnarray}
\begin{figure}[h]
	\centering
	\includegraphics[width=0.9\textwidth]{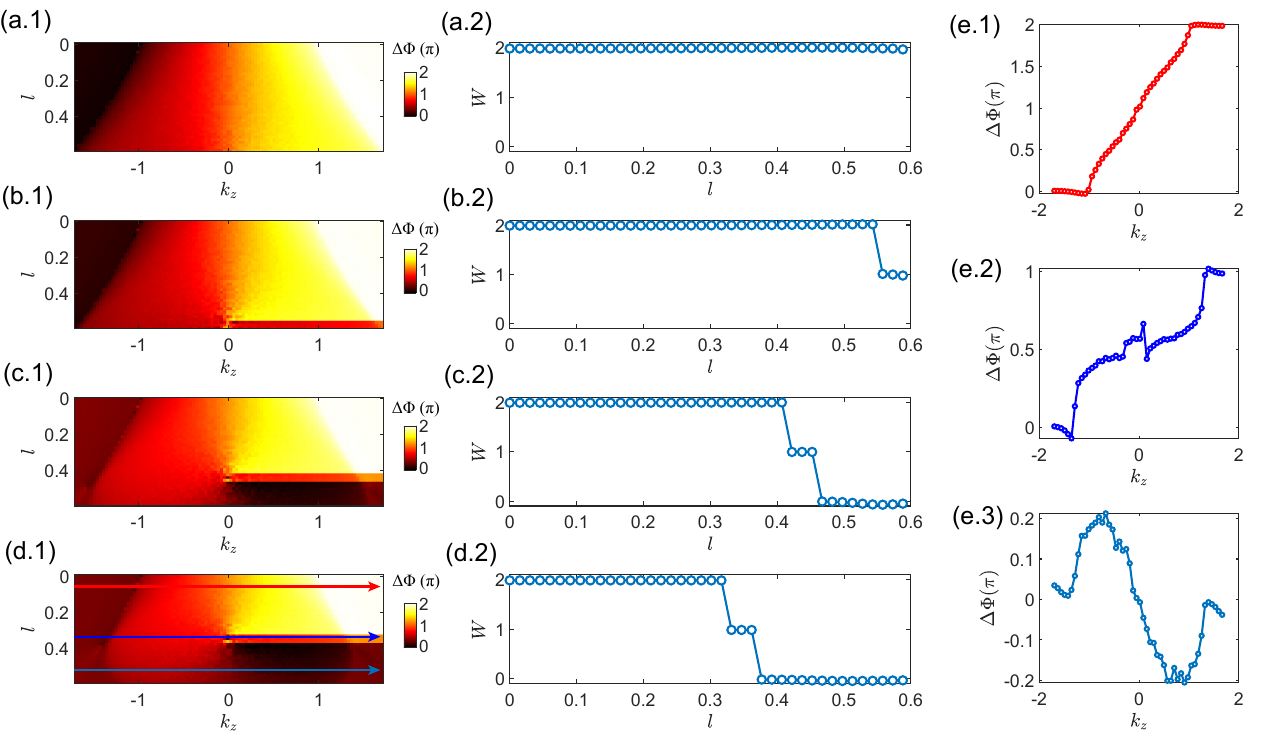}
	\caption{Numerical results of the Wilson loop integral. The nodal-knot phase is the Hopf-linked phase with disorder type $\Delta_z$, which is the same as Fig. 3 (c) and (d) in the main text, with $\Delta_z(0)=0.12$ $, 0.24, 0.36, 0.48$ from (a.1,2) to (d.1,2). The change of the accumulated Berry phase $\Delta\Phi$ with respect to $k_z$ (which labels the slices as depicted in Fig. \ref{figRwilson} (c)) from $k_z=-1.7$ to $k_z=1.7$ and the renormalization scale $l$ are shown in (a.1)-(d.1). The changes of Wilson loop integrals with respect to the renormalization scale $l$ are shown in (a.2)-(d.2). Before the transition, the phase with $W=2$ represents the Hopf-linked nodal-knot. After transtion, $W=0$ represents the unlinked phase. The variation of $\Delta\Phi$ with respect to $k_z$ in (e.1)-(e.3) for three specific $l$ as labeled in (d.1).}
	\label{figRwilsonhopf}
\end{figure}
\begin{figure}[h]
	\centering
	\includegraphics[width=0.9\textwidth]{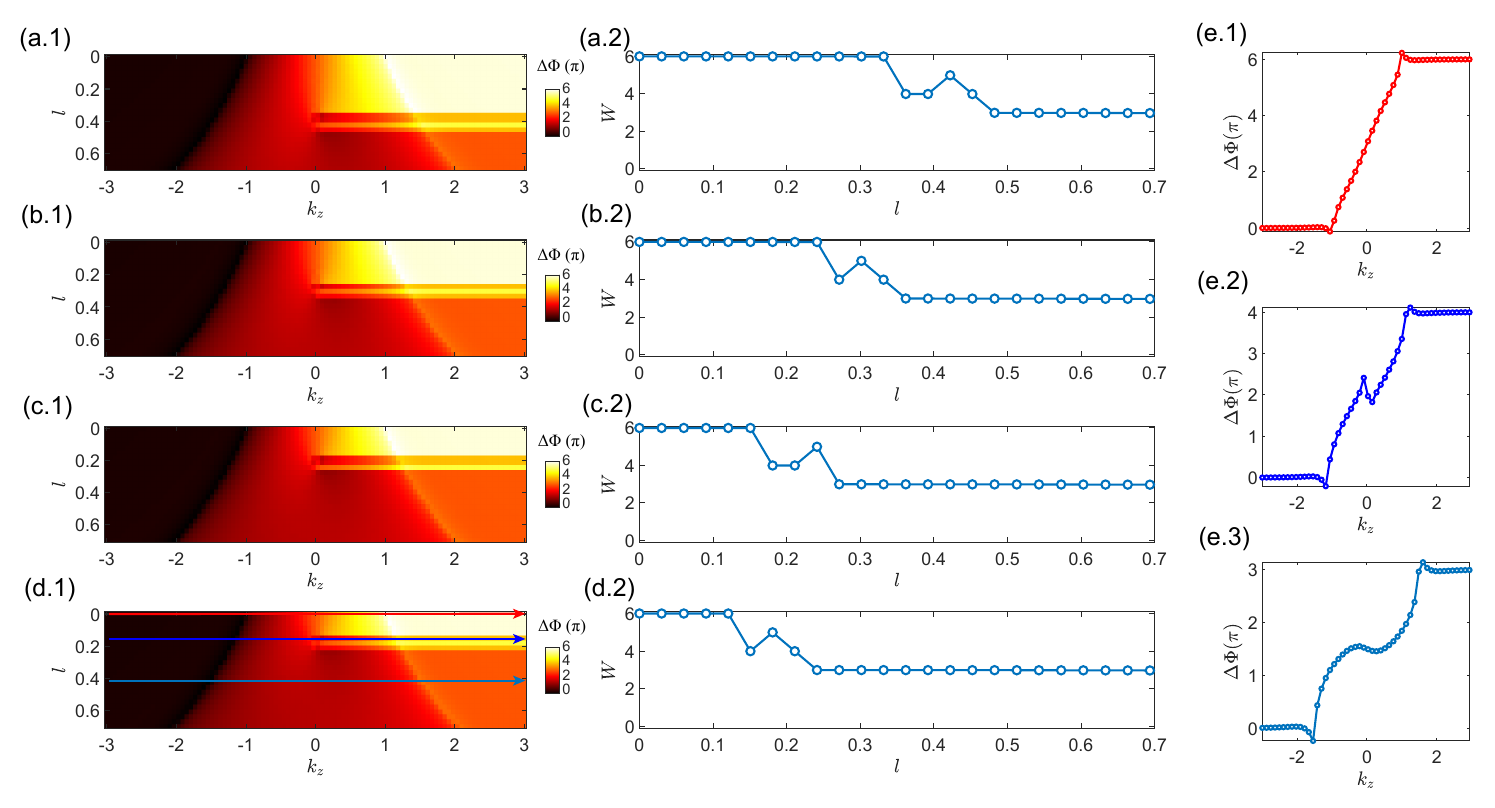}
	\caption{Numerical results of the Wilson loop integral. The nodal-knot phase is the valknut phase with disorder type $\Delta_z$, which is the same as Fig. 4 (c) and (d) in the main text, with $\Delta_z(0)=0.3$ $, 0.5, 0.7, 0.9$ from (a.1,2) to (d.1,2). The change of the accumulated Berry phase $\Delta\Phi$ with respect to $k_z$ (which labels the slices as depicted in Fig. \ref{figRwilson} (c)) from $k_z=-3$ to $k_z=3$ and the renormalization scale $l$ are shown in (a.1)-(d.1). The changes of Wilson loop integrals with respect to the renormalization scale $l$ are shown in (a.2)-(d.2). Before the transition, the phase with $W=6$ represents the valknut knot. After transtion, $W=3$ represents the trefoil knot phase. The variation of $\Delta\Phi$ with respect to $k_z$ in (e.1)-(e.3) for three specific $l$ as labeled in (d.1).}
	\label{figRwilsonvalknut}
\end{figure}
It yields the accumulated Berry phase along the Wilson loop, with $\bm{A}(\bm{k})=-i\langle u_{\bm k}|\partial_{\bm k}|u_{\bm k}\rangle$ the Berry connection. Here, we consider the Wilson loop to be the nodal lines, and the orientation of the integral path $\bm l$ is determined by the tangent vector $\bm{T}(\bm k_0)$ in Eq. (4) in the main text. We should also note that $\bm{A}(\bm{k})$ on the degenerate nodal lines are not well-defined, the Wilson loop integral should be slightly deviate from the nodal lines as depicted in Fig. \ref{figRwilson} (b), and this procedure is sometimes called to ``frame'' the knots into strands (of which the idea was firstly proposed in Witten's paper \cite{witten_quantum_1989}). Therefore, we frame the nodal-knot slightly outward from the origin such that unknotted loop has no contribution to $W$. To obtain the Berry phase along the Wilson loop, we slice 3D Brillouin zone into many 2D layers as schematically numered as 1-5 in Fig. \ref{figRwilson} (c). We view the directed Wilson loop as the worldline of quasi particles that intersect each layer at four points as shown by colored circles in (d), and the number in each circle denotes which layer the worldline intersects. Moreover, the colors (red or blue) of the circles denote the direction of the worldline that pass the corresponding layer. The Wilson loop integral can then be obtained by counting the net phase changes of the intersection points. Speficically, as demonstrated in Fig. \ref{figRwilson} (d), as the slices go upward, the intersection points marked red (blue) between slice and the upward (downward) worldlines move in the slice plane. The upward going worldlines give the net positive Berry phase accumulation denoted by $\Delta \Phi_{\rm red}$, and the downward going worldlines give the net negative Berry phase accumulation denoted by $\Delta \Phi_{\rm blue}$. The total Berry phase alone the worldline (thus the Wilson loop integral) is 
\begin{eqnarray}
\Delta \Phi=\Delta\Phi_{\rm red, 1\to 5}-\Delta\Phi_{\rm blue, 1\to 5}=\pi W(L_1,L_2).
\label{deltaphi}
\end{eqnarray}
Here, $1\to 5$ means the five representative slices shown in \ref{figRwilson} (c), and $L_1$ and $L_2$ denote the two loops that form the Hopf link.

In Fig. \ref{figRwilsonhopf}, we show the numerical results of the Wilson loop integral. The nodal-knot phase is the Hopf-linked phase with disorder type $\Delta_z$, which is the same as Fig. 3 (c) and (d), with $\Delta_z(0)=0.12, 0.24, 0.36, 0.48$ for Fig. \ref{figRwilsonhopf} from (a.1,2) to (d.1,2). Specifically, the changes of Wilson loop integrals with respect to the renormalization scale $l$ as shown in Fig. \ref{figRwilsonhopf} (a.2)-(d.2) is reflected in the change of the accumulated Berry phase $\Delta\Phi$ (Eq. (\ref{deltaphi})) with respect to $k_z$ (which labels the slices as depicted in Fig. \ref{figRwilson} (c)) from $k_z=-1.7$ to $k_z=1.7$ shown in Fig. \ref{figRwilsonhopf} (a.1)-(d.1). Clearly, as $\Delta_z(0)$ increases from 0.12 to 0.48, a transition in $\Delta\Phi$ arises as shown in Fig. \ref{figRwilsonhopf} (c.1) and (d.1). Moreover, the corresponding $\Delta\Phi$ at $k_z=1.7$, which is also the total Wilson loop integral $W=\pi\Delta\Phi$, is shown in Fig. \ref{figRwilsonhopf} (a.2)-(d.2). Before the transition, the phase with $W=2$ represents the Hopf-linked nodal-knot. After transtion, $W=0$ represents the unlinked phase. We also show the variation of $\Delta\Phi$ with respect to $k_z$ in Fig. \ref{figRwilsonhopf} (e.1)-(e.3) for three specific $l$ as labeled Fig. \ref{figRwilsonhopf} (d.1). From the plots we can clearly see that $\Delta\Phi$ increases from 0 to $2\pi$ for the Hopf link while go back to 0 for the unlink. Similarly, numerical results of the Wilson loop integral for the valknut phase are shown in Fig. \ref{figRwilsonvalknut}. The disorder type is $\Delta_z$, which is the same as Fig. 4 (c) and (d) in the main text with $\Delta_z(0)=0.3$ $, 0.5, 0.7, 0.9$ from Fig. \ref{figRwilsonvalknut} (a.1,2) to (d.1,2). The change of the accumulated Berry phase $\Delta\Phi$ with respect to $k_z$ from $k_z=-3$ to $k_z=3$ and the renormalization scale $l$ are shown in Fig. \ref{figRwilsonvalknut} (a.1)-(d.1). The changes of Wilson loop integrals with respect to the renormalization scale $l$ are shown in Fig. \ref{figRwilsonvalknut} (a.2)-(d.2). Before the transition, the phase with $W=6$ represents the valknut knot. After transtion, $W=3$ represents the trefoil knot phase. The variation of $\Delta\Phi$ with respect to $k_z$ in Fig. \ref{figRwilsonvalknut} (e.1)-(e.3) for three specific $l$ as labeled in Fig. \ref{figRwilsonvalknut} (d.1).

By studying both the transitions in $W$ and the singular behavior of the DOS marker near the knot transition points, the boundaries of the nodal-knot transitions can be determined more clearly. Moreover, the knot Wilson loop integral $W$ indeed has real physical meaning and is not merely a mathematical quantity without observable effects. It can represent the total phase shifts for closed cyclotron orbits in the de Haas-van Alphen oscillation experiment.

		\bibliography{nodal_link.bib}